\newif\ifpreprint
  \preprinttrue		

\ifpreprint
	\documentstyle[aas2pp4,epsf,twoside,fleqn]{article}
\else
	\documentstyle[12pt,aasms4,epsf]{article}
\fi

\newcommand{\etal}	{{\it et al.}}
\newcommand{\hi}	{\mbox{H\,{\sc i}}}
\newcommand{\hip}	{HIPPARCOS}

\newcommand{\paI}	{\mbox{\rm paper\,I}}
\newcommand{\saI}	{\mbox{\rm Sample\,1}}

\newcommand{\tsy}	{\textstyle}
\newcommand{\bea}	{\begin{array}}
\newcommand{\eea}	{\end{array}}
\newcommand{\beq}	{\begin{equation}}
\newcommand{\eeq}	{\end{equation}}
\newcommand{\ben}	{\begin{eqnarray}}
\newcommand{\een}	{\end{eqnarray}}

\newcommand{\B}[1]	{\mbox{\boldmath$ #1 $}}
\newcommand{\bnabla}	{{\B{\nabla}}}
\newcommand{\bx}        {\B{x}}
\newcommand{\bvel}      {\B{v}}
\newcommand{\br}	{\B{r}}
\newcommand{\hbr}	{\hat{\br}}
\newcommand{\bp}        {\B{p}}
\newcommand{\bq}        {\B{q}}
\newcommand{\bkappa}    {\B{\kappa}}
\newcommand{\bk}        {\B{k}}
\newcommand{\bl}        {{\B{l}}}

\newcommand{\bsl}       {{\B{\scriptstyle l}}}
\newcommand{\bsn}       {{\B{\scriptstyle n}}}
\newcommand{\by}        {\B{y}}
\newcommand{\bA}        {{\bf A}}
\newcommand{\bR}        {{\bf R}}
\newcommand{\bS}        {{\bf S}}
\newcommand{\D}		{{\rm d}}
\newcommand{\cQ}	{\mbox{${\cal Q}_\alpha$}}
\newcommand{\tcQ}	{\mbox{$\tilde{\cal Q}_\alpha$}}
\newcommand{\hf}	{\mbox{$\hat{f}$}}
\newcommand{\hfa}	{\mbox{$\hat{f}_\alpha$}}
\newcommand{\hphi}	{\mbox{$\hat{\phi}$}}
\newcommand{\cA}	{\mbox{$\cal A$}}
\newcommand{\tcA}	{\mbox{$\tilde{\cal A}$}}
\newcommand{\cS}	{\mbox{$\cal S$}}
\newcommand{\cL}	{\mbox{$\cal L$}}
\newcommand{\cN}	{\mbox{$\cal N$}}
\newcommand{\p}		{\partial}
\newcommand{\ave}[1]    {\left\langle{#1}\right\rangle}

\newcommand{\mkms}	{{\rm km}\,{\rm s}^{-1}}
\newcommand{\kms}	{\mbox{$\,{\rm km}\,{\rm s}^{-1}$}}
\newcommand{\kpc}	{\mbox{$\,{\rm kpc}$}}
\newcommand{\kmskpc}	{\mbox{$\,{\rm km}\,{\rm s}^{-1}\,{\rm kpc}^{-1}$}}
\newcommand{\parc}	{\mbox{$\,{\rm pc}$}}

\newcommand{\magn}	{\mbox{\,{\rm mag}}}

\ifpreprint
\else
  \lefthead{Walter Dehnen}
  \righthead{The distribution of nearby stars in velocity space}
\fi
\slugcomment{accepted for publication by the Astronomical Journal}

\begin{document}

\ifpreprint \thispagestyle{empty} \fi
\title{The Distribution of Nearby Stars in Velocity Space \\
       Inferred from Hipparcos Data}
\author{Walter Dehnen}
\affil{Theoretical Physics, 1 Keble Road, Oxford OX1 3NP, United Kingdom}

\begin{abstract}
The velocity distribution $f(\bvel)$ of nearby stars is estimated, via a 
maximum-likelihood algorithm, from the positions and tangential velocities of a
kinematically unbiased sample of 14\,369 stars observed by the \hip\ satellite.
$f$ shows rich structure in the radial and azimuthal motions, $v_R$ and 
$v_\varphi$, but not in the vertical velocity, $v_z$: there are four prominent 
and many smaller maxima, many of which correspond to well known moving groups. 
While samples of early-type stars are dominated by these maxima, also up to 
about a quarter of red main-sequence stars are associated with them. These 
moving groups are responsible for the vertex deviation measured even for 
samples of late-type stars; they appear more frequently for ever redder samples;
and as a whole they follow an asymmetric-drift relation, in the sense that 
those only present in red samples predominantly have large $|v_R|$ and lag in 
$v_\varphi$ w.r.t.\ the local standard of rest (LSR). 
The question arise, how these old moving groups got on their eccentric orbits?
A plausible mechanism known from the solar system dynamics which is able to 
manage a shift in orbit space is sketched. This mechanism involves locking into
an orbital resonance; in this respect is intriguing that Oort's constants, as 
derived from \hip\ data, imply a frequency ratio between azimuthal and radial 
motion of exactly $\Omega:\kappa=3:4$.
 
Apart from these moving groups, there is a smooth background distribution,
akin to Schwarzschild's ellipsoidal model, with axis ratios $\sigma_R:
\sigma_\varphi:\sigma_z\approx1:0.6:0.35$. The contours are aligned with the 
$v_r$ direction, but not w.r.t.\ the $v_\varphi$ and $v_z$ axes: the mean $v_z$ 
increases for stars rotating faster than the LSR. This effect can be explained
by the stellar warp of the Galactic disk. If this explanation is correct, 
the warp's inner edge must not be within the solar circle, while its pattern 
rotates with frequency $\gtrsim13\kmskpc$ retrograde w.r.t.\ the stellar orbits.
\end{abstract}


\ifpreprint
  \section{INTRODUCTION}
\else
  \section{Introduction}
\fi
The dynamical state of a stellar system is completely described by its 
phase-space distribution function $F(\bx,\bvel)$. Knowledge of that function 
allows one to derive the underlying gravitational potential and hence the mass
distribution, since in equilibrium $F$ must be constant on the orbits supported
by the potential. Furthermore, one can try to understand $F$ in terms of the 
possible formation history and evolution of the stellar system under study. 
Unfortunately, $F$, being a six-dimensional function, is impossible to measure 
for any remote galaxy; all one can hope is to measure a three-dimensional 
projection, and even this requires spectra to be taken at all positions on the 
galaxy. For the Milky Way the situation is completely different. In a sense it 
is worse, since we cannot observe a clean and easily predictable projection. 
Rather we observe individual stars and are in danger of seeing the trees for the
wood. However, if we forget individual stars and observe large samples, we can 
actually measure $F(\bx,\bvel)$ itself, at least in principle. The problem here 
is the shear amount of labour necessary to collect accurate measurements of 
stellar phase-space coordinates for large samples, in particular of faint stars.
This has so far prevented this route from being used. Thus, we must either 
restrict ourselves to luminous and hence young stars, or to very nearby stars 
of all stellar types and ages. Since young stars are more likely to have not 
yet settled into equilibrium, interpretation of their phase-space distribution 
is more complicated. The best we can currently hope to get for the old stellar 
population is the velocity distribution in the solar neighborhood $f(\bvel)
\equiv F(\bx{=}\bx_\odot,\bvel)$.

Historically, there are two divergent approaches to $f(\bvel)$. One is mainly
based on theoretical lines of argument like the epicycle approximation and 
dynamical heating mechanisms. It manifests itself in Schwarzschild's (1908) 
ellipsoidal distribution, i.e.\ $f(\bvel)$ is assumed to be smooth, 
single-peaked, and well described by the mean velocity and dispersion in 
conjunction with Str\"omberg's asymmetric drift relation (cf.\ Binney \& 
Tremaine 1987, p.~198 and 202). The other approach to $f(\bvel)$ is more 
observational and mainly restricted to luminous stars on account of magnitude 
limits. It begins with Kapteyn's (1905) recognition of moving groups, which 
were later studied in more detail by Eggen (1965 and references therein, 1995, 
1996). That is, $f(\bvel)$ for early-type stars appears to be dominated 
by several independent components. According to the standard interpretation, 
stars of a moving group have formed simultaneously in a small phase-space 
volume. The idea is that phase mixing and scattering processes have not yet 
completely washed out the initial correlation of the moving group, and we can 
still observe a stream of stars with similar velocities. The old stellar 
populations, on the other hand, should be completely mixed and obey a smooth 
distribution function. This interpretation, however, is merely hypothetical and 
by no means proven. Indeed, the results of this paper raise problems for this
simple picture.

ESA's astrometric satellite \hip\ (\cite{hipcat}) provided us with positions and
trigonometric parallaxes of unprecedented accuracy for tens of thousands of 
stars near the Sun. From these data we can, for the first time in history, 
extract large and homogeneous stellar samples with accurately known distances 
and, what is important, completely free of the kinematic biases that have 
plagued similar studies in the past. Since \hip\ has additionally measured the 
absolute proper motions for the same stars, it offers us the opportunity to 
investigate in some detail the nature of the velocity distribution in the solar
neighborhood not only for early-type stars but also for the old stellar 
population of the Galactic disk. In a preceding paper (Dehnen \& Binney 1998b, 
hereafter \paI), we have used a kinematically unbiased subsample of the \hip\ 
Catalogue to infer, as a function of stellar color, the first two moments, mean
and dispersion, of $f(\bvel)$ for nearby stars. This paper goes beyond these 
moments to infer the velocity distribution itself from the \hip\ data.

If the \hip\ mission were complemented by a program to measure the radial
velocities of the same stars, we could obtain to high accuracy their space 
velocities $\bvel$, and hence directly measure the distribution function 
for these stars. Unfortunately, however, the radial velocities for many stars in
the \hip\ Catalogue are not (yet) publicly known. From the existing literature 
one can extract radial velocities only for a subset of \hip\ stars that is 
heavily kinematically biased in the sense that it predominantly contains 
high-proper-motion stars (\cite{Venice1}). Hence, in order to make full use of 
the \hip\ Catalogue, or of similar future catalogs, for studies of stellar 
kinematics, one cannot necessarily rely on radial velocities, but must be able 
to work with the positions and tangential velocities alone. For each position on
the sky, the tangential velocity is a certain combination of the components of 
the space velocity. It is immediately clear that we can only proceed if we link 
together the tangential velocities measured for stars in different directions 
from the Sun. In order for such a method to be valid, one must make the basic 
assumption that $f(\bvel)$ is independent of the position on the sky, a 
condition that should be satisfied for sufficiently nearby stars. 

In Section~2, I show that, under this basic assumption, knowing the 
distributions of tangential velocities for a region of the celestial sphere 
that cuts all great circles gives complete knowledge of the full distribution
of space velocities. The problem is formally equivalent to that of classical 
tomography. However, there is a fundamental difference, namely that we are 
unable to measure the distributions of tangential velocities to arbitrary 
accuracy, rather we only know the tangential velocities for a few thousand 
stars distributed over the sky. Consequently, we cannot use the tools developed
for tomography. Instead, in Section~3, a maximum-likelihood technique is 
presented and tested that enables one to estimate $f(\bvel)$ for a set of 
stars from their positions, parallaxes, and proper motions. 

The reader not interested in these techniques and numerical details but mainly 
in the resulting $f(\bvel)$ may skip Sections~2 and 3 and directly go to 
Section~4, where the algorithm is applied to the kinematically unbiased 
subsample of \hip\ stars that we have already used in \paI. The results and 
their implications are discussed in Sections~5 and 6, while Section~7 sums up.

\ifpreprint
  \section{THE PROJECTION OF SPACE VELOCITIES} 
\else
  \section{The Projection of Space Velocities} 
\fi
\subsection{The Projection Equations} \label{sec-proj-eq}
Let $\bvel$ be the velocity of a star w.r.t.\ the Sun in a Cartesian
coordinate system with $v_x$, $v_y$, and $v_z$ denoting the motions towards the
Galactic center, in the direction of Galactic rotation, and towards the north
Galactic pole, respectively. While this coordinate system is a good choice in
dynamical studies, the natural coordinates for observations are $v_r=\dot{r}$,
$v_\ell=r\mu_\ell\cos b$, and $v_b=r\mu_b$. Here $\ell$, $b$, and $r$ 
denote, respectively, Galactic longitude, latitude, and distance from the 
Sun. $\mu_\ell$ and $\mu_b$ are the proper motions corrected for the 
effects of Galactic rotation
\beq \label{pm-corr}
 \bea{r@{\,=\,}l} 
 \mu_\ell & \mu_\ell^{\rm(obs)}  -  A \, \cos(2\ell) - B \\ 
 \mu_b    & \mu_b^{\rm(obs)}     +  A \, \sin(2\ell) \, \cos b \, \sin b,
\eea \eeq 
with Oort's constants $A$ and $B$; I use Feast \& Whitelock's (1997) values
(in \kmskpc) $A=14.82$ and $B=-12.37$ obtained from \hip\
Cepheids, but the results are insensitive to the precise values. The two 
coordinate systems are related by a rotation
 \beq (v_r,v_\ell,v_b)^T = \bR\cdot\bvel, \qquad  
	\bvel = \bR^T\!\cdot(v_r,v_\ell,v_b)^T \eeq
with rotation matrix
 \beq	\bR = \left[\bea{r@{\quad}r@{\quad}r}
         \cos b\,\cos\ell &  \cos b\,\sin\ell & \sin b \\
        -\sin \ell        &  \cos \ell        & 0      \\ 
        -\sin b\,\cos\ell & -\sin b\,\sin\ell & \cos b \eea \right]. \eeq
For our purposes it is useful to consider the two-dimen\-sio\-nal velocity 
vector $\bq\equiv(v_\ell,v_b)$, which can be evaluated from \hip's measurements.
It is related to $\bvel$ by the projection equation
 \beq\label{proj-one}
        \bq = \bS\cdot\bvel
 \eeq
where the $2\times3$ matrix $\bS$ is given by the second and third row of
$\bR$. In the $(v_x,v_y,v_z)$ frame, the corresponding tangential velocity
$\bp\equiv\bS^T\cdot\bq$ is related to $\bvel$ by the projection
\beq \label{proj-two}
        \bp=\bvel-(\hbr\cdot\bvel)\hbr\equiv\bA\cdot\bvel,
\eeq
where $\hbr$, given by the first row of $\bR$, is the unit vector
in the direction ($\ell,b)$. Equation (\ref{proj-two}), which is identical
to Equation (3) of \paI, defines the projection matrix $\bA={\bf I}-\hbr
\otimes\hbr$ (\paI, equation 4) and is equivalent to (\ref{proj-one}).
Clearly, matrix $\bA$ is singular, so (\ref{proj-two}) cannot be 
inverted; one needs the radial velocity to obtain $\bvel$. However, taking
the average of (\ref{proj-two}) over a region of the sky, which can in principle
be as small as a line segment, results in the non-singular equation
 \beq 
	\ave{\bp} = \ave{\bA} \cdot \ave{\bvel},
 \eeq
which can be solved for the average space velocity $\langle\bvel\rangle$, as we
have done in \paI. Similarly, one can take the average of the $n$th outer 
product of (\ref{proj-two}) with itself
 \beq 	\ave{\bp\otimes\dots\otimes\bp} =
      	\ave{\bA\otimes\dots\otimes\bA} \cdot 
	\ave{\bvel\otimes\dots\otimes\bvel}
 \eeq
and obtain the moments of order $n$ for the space velocities; in \paI\ we 
have done this for the velocity dispersion tensor 
$\B{\sigma}^2\equiv\ave{\bvel\otimes\bvel}-\ave{\bvel}\otimes\ave{\bvel}$.

The projection onto the celestial sphere can also be formulated in terms of the
velocity distribution function $f(\bvel)$. To this end let $\rho(\bq|\hbr)$ be
the probability distribution of tangential velocities in direction $\hbr$, then 
\beq \label{proj-thr}
	\rho(\bq|\hbr) = \int \D v_r\,
		   f\left(\bvel = \bS^T\!\cdot\bq + v_r \hbr\right).\\
\eeq

\subsection{The Fourier Slice Theorem}
At this point is very fruitful to invoke the Fourier slice theorem. This follows
directly by Fourier transforming (\ref{proj-thr}) and states that the 
(two-dimensional) Fourier transform $\varrho(\bkappa|\hbr)$ of $\rho(\bq|\hbr)$
w.r.t.\ $\bq$ is related to the Fourier transform ${\cal F}(\bk)$ of 
$f(\bvel)$ by
\beq \label{eq-fst}
	\varrho(\bkappa|\hbr) = \sqrt{2\pi}\,{\cal F}(\bk=\bS^T\cdot\bkappa),
\eeq
i.e.\ $\varrho$ is given by ${\cal F}$ in the slice normal to $\hbr$. Thus, one 
will have full knowledge of ${\cal F}(\bk)$, and hence of $f(\bvel)$, if and 
only if one knows $\rho(\bq|\hbr)$ for a set of $\hbr$ such that the 
corresponding set of slices covers the full $\bk$ space. This immediately leads
to the following statement:

\medskip
\noindent
``The underlying $f(\bvel)$ is uniquely determined by its projection $\rho(\bq
|\hbr)$ if and only if the latter is known for a region of the celestial sphere
that intersects with every great circle.''

\medskip
\noindent
The smallest such region is any half great circle. Hence, knowing $\rho(\bq|\hbr
)$ for more than that yields redundant information that, in principle, can be 
used to check our basic assumption that the inferred $f(\bvel)$ does not 
depend on direction from the Sun.

The attentive reader might feel an apparent contradiction between the statement
above and the result from Section~\ref{sec-proj-eq} that the moments of 
$f(\bvel)$ can be derived already if one has precise knowledge of $\rho(\bq|
\hbr)$ on a small region of the sky. The resolution is as follows. Given one 
knows $\rho(\bq|\hbr)$ on a region that does not satisfy the above criterion, 
then there will be a region in $\bk$ space where one has no knowledge of ${\cal
F}(\bk)$. Hence, any distribution $f^\prime(\bvel)$ whose Fourier transform 
${\cal F}^\prime(\bk)$ is non-zero only in this zone of ignorance is 
undetectable in the data (it projects to zero). Such a function cannot possibly
be analytic (in the mathematical sense), i.e.\ the Taylor series for 
${\cal F}^\prime(\bk)$ does not converge globally. Since
\[
        \int\!\D^3\!\bvel f(\bvel)\, v_x^lv_y^mv_z^n
        = {\partial^{l+n+m} {\cal F}\over
           \partial (ik_{v_x})^l\,
           \partial (ik_{v_y})^m\,
           \partial (ik_{v_z})^n } \bigg|_{\bk=0},
\]
we can only infer ${\cal F}(\bk)$ from the moments of $f(\bvel)$ if we assume
that ${\cal F}$ is analytic. The Fourier slice theorem makes no assumptions 
about the nature of $f$, and, as a consequence, stronger observational 
constraints are required. 

\subsection{The Marginal Distributions}\label{sec-proj-marg}
In many cases one is only interested in the distribution of velocities in the 
plane $f_\parallel(v_x,v_y) \equiv \int\D v_z\,f(\bvel)$. A projection 
equation for $f_\parallel$ can be obtained by integrating (\ref{proj-thr}) over
all $v_b$:
\beq \label{proj-par}
	\rho(v_\ell|\ell) = \int\D v_\parallel\;f_\parallel
		\left({v_x = \cos\!\ell\,v_\parallel-\sin\!\ell\,v_\ell \atop
	 	       v_y = \sin\!\ell\,v_\parallel+\cos\!\ell\,v_\ell}\right),
\eeq
where $\rho(v_\ell|\ell)$ is the distribution of $v_\ell$ for stars in direction
$\ell$. That is, we could derive $f_\parallel$ from the $v_\ell$ alone, without
need of $v_b$, and still use data from stars all over the sky. However, as
experiments showed, inverting (\ref{proj-thr}) and then projecting gives better
results than inverting (\ref{proj-par}). This is not surprising, since the 
star's latitudinal velocities contain valuable information about $(v_x,v_y)$.

For the recovery of the marginal distribution of vertical velocities,
$f_\perp(v_z)\equiv\int\!\!\!\int\D v_x\D v_y\,f(\bvel)$,
directly from the data we can only use stars in the Galactic plane. This follows
from the Fourier slice theorem: the only slices that give information on the 
$k_z$ axis are those that contain it, i.e.\ $\rho(v_b|b{=}0)=f_\perp(v_b)$. 

\subsection{Relation to Classical Tomography}
There is a close connection between our problem and tomography as in medical
diagnostic. In that case, the unknown distribution is defined in configuration 
rather than in velocity space, but the projection equation is identical to 
(\ref{proj-thr}) or (\ref{proj-par}). There is, however, a fundamental 
difference to our problem: astronomy relies on {\em observations\/} -- most 
other sciences rely on {\em experiments}. In tomography one is able to measure 
the projected distribution to high accuracy for many directions densely
covering half a great circle. A viable way to recover the underlying 
distribution from the density of $\ga10^{11}$ gathered (or absorbed) photons
is then the direct inversion via the Fourier slice theorem, also known as 
inverse radon transform. By contrast, we only know, typically, $\sim10^4$ 
points that nature has randomly chosen for us from the underlying $f(\bvel)$. 
The great gulf between $10^{11}$ and $10^4$ obliges us to analyze our data 
differently.

\ifpreprint
  \section[]{THE VELOCITY DISTRIBUTION AS  \\
	   MAXIMUM-PENALIZED-LIKELIHOOD- \\
	   ESTIMATE}
\else
  \section{The Velocity Distribution as Maximum-Penalized-Likelihood-Estimate}
\fi
\label{sec-alg}
This section deals with the algorithmic and numerical aspects of estimating
$f(\bvel)$ from the tangential velocities and directions of $N$ stars. In order
to avoid confusion, I will use, in this section only, the symbol $f$ for any 
model or estimate of the true and unknown distribution $f_0$ that underlies the
data.

\subsection{Formulation}
As the discussion in the last subsection showed, it is rather hopeless to infer
$f_0(\bvel)$ directly via the Fourier slice theorem. Rather, following Bayesian
statistics, the route to estimate $f_0(\bvel)$ is to maximize the log-likelihood
of a model\footnote{
	In the literature one can also find convergent point methods used for 
	similar purposes. These methods rely on geometrical arguments and stem
	from the times when more rigorous treatment (like maximizing the 
	log-likelihood) was impractical on technical grounds. Nonetheless, 
	Chereul, Cr\'ez\'e \& Bienaym\'e (1997) have used such a technique for 
	the estimation of $f_0(\bvel)$ for A stars from \hip\ data. These 
	authors define the convergent points for a pair of stars as the 
	velocities on their lines $\bvel=\bp+v_r\hbr$ (where the unknown $v_r$ 
	is treated like a variable) that minimize the distance between them. If
	this distance is smaller than some threshold, the two velocities are 
	remembered. After considering all possible pairs of stars, this results 
	in a list of velocities, that can be directly translated into a 
	histogram to estimate $f_0(\bvel)$. Testing this method for anisotropic
	$f_0(\bvel)$, I found that the outer contours are always too round and 
	the estimated velocity dispersions significantly too high. This failure 
	can be attributed to accidental convergent points at large $|\bvel|$. 
	In the limit of infinite many stars, this form of the convergent-point
	method does not converge to $f_0$ -- in contrast to a 
	maximum-likelihood technique.}
$f(\bvel)$
\beq
        \cL(f) = N^{-1}\; \sum_{k=1}^N\; \ln P(\bq_k|\hbr_k,f),
\eeq
where
\ben \label{prob-k}
        P(\bq_k|\hbr_k,f)&=&\int \D v_r\;f(\bvel=\bp_k+v_r\hbr_k)\\
     \label{prob-k-delta}
		&=& \int \D^3\!\bvel\; f(\bvel)\;\delta(\bq_k-\bS_k\cdot\bvel)
\een
is the probability to observe the tangential velocity $\bp_k=\bS_k^T\cdot\bq_k$ 
for a star seen in direction $\hbr_k$ and with velocity drawn from $f(\bvel)$%
\footnote{
	The measurement uncertainties of the \hip\ data can be incorporated
	by replacing the $\delta$-function in (\ref{prob-k-delta}) with a 
	Gaussian in the observables. However, as became clear in \paI, the
	dominant source of errors is sampling noise due to the small number 
	of stars in the sample, and accounting for the uncertainties would 
	hardly improve the results.}. 
The functional $\cL(f)$, however, has no maximum, it is unbounded, because one
can arrange $N$ $\delta$-functions, one on each line $\bp_k+v_r\hbr_k$ in 
$\bvel$ space, to get $\cL=\infty$. In order to obtain a well defined procedure,
one needs to regularize $\cL(f)$. This is conveniently done by adding a penalty 
functional, which in Bayesian statistic might be interpreted as the logarithm 
of the prior. That is, one maximizes
\beq
        \cQ(f) = \cL(f) - {\textstyle {1\over 2}} \alpha \cS(f),
\eeq
where the penalty functional $\cS(f)$ measures the roughness of $f(\bvel)$, 
whereas $\alpha>0$ is a Lagrange multiplier, usually referred to as 
smoothing parameter. The function that maximizes $\cQ(f)$ subject to the 
constraints
\ben    \label{eq-pos} 0 &\le& f(\bvel) \quad{\rm and}\\ 
        \label{eq-norm} 1 &=&   \cN(f) \equiv \int \D^3\bvel\; f(\bvel)
\een
is the maximum-penalized-likelihood-estimate (MPLE) and will
be denoted $\hfa(\bvel)$. The optimal choice of $\alpha$ is a problem on its 
own and the subject of the next subsection. A common choice for the penalty 
functional is $\int\D^3\bvel\,(\bnabla^2\ln\! f)^2$. I will use a slightly 
different form:
\beq \label{eq-def-S}
        \cS(f)=\int\D^3\bvel\,(\tilde{\bnabla}^2 \ln\! f)^2
\eeq
with 
\beq \label{eq-def-nablatilde}
        \tilde{\bnabla}\equiv \left(\tilde{\sigma}_x {\p\over\p v_x},\,
                                    \tilde{\sigma}_y {\p\over\p v_y},\,
                                    \tilde{\sigma}_z {\p\over\p v_z}   \right)
\eeq
where $\tilde{\sigma}_i$ is a measure for the width of $f(\bvel)$ in the $i$th 
dimension, for instance, the velocity dispersion estimated via the methods of 
paper~I. With this choice of $\cS(f)$, the smoothing parameter $\alpha$ is
independent of the width of $f_0$.

\subsection{The Optimal Smoothing}\label{sec-aopt}
The parameter $\alpha$ determines the amount of smoothing. Clearly, there 
exists an optimum value, $0<\alpha_{\rm opt}<\infty$, since neither $\alpha=0$ 
(no smoothing leading to the $\delta$-function catastrophe) nor $\alpha=\infty$
(ignoring the data) gives an useful result. One usually finds $\alpha_{\rm opt}$
by requiring that it minimizes the difference, $D(\hfa,f_0)$, between $\hfa$ 
and $f_0$. There are various ways to measure this difference, most commonly 
used are (cf.\ Silverman 1986) the integrated square error (ISE)
\beq\label{eq-ise}
	D(\hfa,f_0) = \int\D^3\!\bvel\;(\hfa-f_0)^2,
\eeq
or the Kullback-Leiber information-distance (KLD)
\beq\label{eq-kld}
        D(\hfa,f_0) = - \int\D^3\!\bvel\; f_0\,\ln(f_0/\hfa).
\eeq
Of course, since $f_0$ is unknown, the ISE (or KLD) cannot be evaluated. We can,
however, estimate $f_0$ and simulate data and analysis, in order to estimate
$D(\hfa,f_0)$. Since $\hfa$ not only depends on $f_0$ and $\alpha$, but also 
on the particular random realization of the $N$ data, $D(\hfa,f_0)$ is a random
variable. Minimizing a random variable make no sense, and one usually considers
the mean over the data realizations, giving the mean integrated square error 
(MISE) and the mean Kullback-Leiber distance (MKLD), respectively. These can be
estimated and hence minimized, even though $f_0$ is unknown, a process called 
cross-validation.
For the present problem I use the following procedure. Starting with some 
initial guess, $\alpha_0$, for $\alpha_{\rm opt}$ one approximates $f_0$ as
$\hf_{\alpha_0}$ and measures the MISE (or MKLD) that results for a given 
$\alpha$. (This is done by drawing $M$ samples of $N$ pseudo-data from 
$\hf_{\alpha_0}$, evaluating for each of them $\hfa$, measuring $D(\hfa,
\hf_{\alpha_0})$, and taking the mean over the $M$ samples.) The value 
for $\alpha$ that minimizes the MISE (or MKLD) is then used as the next iterate
for $\alpha_{\rm opt}$. If one starts off with a good guess (`by eye'), this 
procedure converges to reasonable accuracy within a few iterations.

Since the data are too poor to infer the full three-dimensional structure of
$f(\bvel)$ ($N$ of $\sim10^4$ corresponds to $\sim30$ stars per dimension), I 
will concentrate on the projections onto the three principal coordinate planes.
In the $v_xv_y$ plane, $f(\bvel)$ is most extended (as judged from the 
dispersions obtained in \paI), and consequently I estimated the MISE and MKLD 
only for the projection onto this plane, i.e.\ instead of the ISE 
(eq.~\ref{eq-ise}) I compute
\beq
	\int\D v_x\,\D v_y \left(  \int\D v_z \left[\hf-f_0\right] \right)^2
\eeq
and analogously for the KLD. 

\subsection{Numerics}
\subsubsection{The MPLE as Unconstrained Extremum}
Our problem is, to find the maximum of $\cQ(f)$ subject to the constraints 
(\ref{eq-pos}) and (\ref{eq-norm}). Constraining the possible solutions in such
a way is awkward, in particular for numerical extremization. Fortunately, the 
non-negativity and normalization conditions can be met without imposing any 
constraint. According to Silverman (1982) the condition $\cN(f)=1$ can be 
satisfied by maximizing instead of $\cQ(f)$ the functional
\beq \label{eq-tcQ-def}
        \tcQ(f) \equiv \cQ(f) - \cN(f).
\eeq
Silverman has shown that the MPLE, $\hfa$, unconditionally maximizes 
$\tcQ(f)$. To see this, let $f^\ast=f/\cN(f)$ be the normalized counterpart of 
$f(\bvel)$. Since $\cS$ only involves derivatives of $\ln\!f$, $\cS(f)=
\cS(f^\ast)$ and elementary manipulations yield
\beq
        \tcQ(f^\ast) = \tcQ(f) + \cN(f) - \ln\cN(f) - 1.
\eeq
Since $t-\ln t-1\ge0$ for all $t\ge0$ with equality only if $t=1$, $\tcQ(f^\ast)
\ge\tcQ(f)$ with equality only if $\cN(f)=1$. Thus, for the maximum of
$\tcQ$, $\cN(f)=1$; but subject to $\cN(f)=1$, $\tcQ(f)$ and $\cQ(f)-1$ are 
identical and the proof is complete.\footnote{
	Note that, even though Silverman's original proof relies on the 
	particular form of the functional to be maximized, one can for any 
	functional $\cA(f)$ ($f\ge0$) eliminate the normalization constraint 
	using his method: the (unconditional) maximum of  
	\beq \label{eq-tcA-def}
		\tcA(f) \equiv \cA(f/\cN[f]) + \ln \cN(f) - \cN(f)
	\eeq
	maximizes $\cA(f)$ subject to $\cN(f)=1$.}

The non-negativity condition (\ref{eq-pos}) can be trivially met by defining
\beq
        f(\bvel) \equiv \exp\left[\phi(\bvel)\right]
\eeq
and considering $\tcQ$ a functional of the unbounded $\phi(\bvel)$. 

\subsubsection{Numerical Representation}
The easiest way to represent $\phi(\bvel)$ numerically is by pixelization on a 
Cartesian grid with $L\equiv L_x\times L_y\times L_z$ cells of size $h_x\times
h_y\times h_z$:
\beq \label{num-rep}
        \phi(\bvel) = \sum_{\bsl} \phi_\bsl\; W_{\bsl}(\bvel),
        \qquad l_i=0,\dots,L_i-1,
\eeq
with the window functions
\beq
        W_{\bsl}(\bvel) = \left\{\bea{l@{\quad}l}
                (h_x h_y h_z)^{-1} 
                        & {\rm if}\quad \forall_i\,|v_i-l_ih_i-y_i|\le 
                                                        {h_i\over 2}, \\
                0       & {\rm otherwise,} \eea \right.
\eeq
where $\by$ is the center of cell $\bl=0$ and determines the position of the
cells in $\bvel$ space. Inserting (\ref{num-rep}) into (\ref{prob-k}) gives
\beq
        P(\bq_k|\hbr_k,f) = \sum_{\bsl}\, {\rm e}^{\tsy\phi_{\bsl}}\,K(k|\bl),
\eeq
where
\beq
        K(k|\bl) \equiv \int\D v_r\;
        W_{\bsl}\left(\bvel=\bp_k + v_r \hbr_k\right),
\eeq
which simply is $(h_xh_yh_z)^{-1}$ times the length in $v_r$ of the segment
of the line $\bvel=\bp_k+v_r\hbr_k$ that lies in cell $\bl$. We can estimate
\beq
        \tilde{\bnabla}^2 \ln f({\bvel}_{\bsl}) \simeq
				\sum_{\bsn} \phi_{\bsn}\;\Xi_{\bsn,\bsl},
\eeq
where
\beq \label{eq-mod-nabla-num}
        \Xi_{\bsn\bsl} = \sum_{i=x,y,z} {\tilde{\sigma}_i^2\over h_i^2} \,
                \left( - 2 \delta_{\bsn,\bsl}
                       + \delta_{\bsn,\bsl+\hat{e}_i}
                       + \delta_{\bsn,\bsl-\hat{e}_i} \right)
\eeq
with $\hat{e}_i$ denoting the unit vector in the $i$th direction.
Thus, the numerical approximation to the functional $\tcQ(\phi)$ is
\ifpreprint \ben \tcQ(\phi) &=& \else \beq \tcQ(\phi) = \fi
		N^{-1} \sum_k \ln\! \bigg[\sum_\bsl\,
		{\rm e}^{\tsy \phi_{\bsl}}\,K(k|\bl]\bigg]
		- \sum_\bsl {\rm e}^{\tsy \phi_\bsl}
\ifpreprint \nonumber\\ & & \fi
		- {\textstyle{1\over2}}\alpha h_xh_yh_z\sum_\bsl
		\bigg[\sum_\bsn\,\phi_\bsn\,\Xi_{\bsn\bsl}\bigg]^2
\ifpreprint \een \else \eeq \fi
with $(\phi_\bsl)\equiv\B{\phi}$, a $L$-dimensional vector.

\subsubsection{Maximization of $\tcQ$}
At the maximum of $\tcQ$, its gradient vanishes giving the relation (with
equation \ref{eq-tcQ-def})
\beq\label{fix-point}
        \phi_\bsl = \ln {\p\cQ\over\p \phi_\bsl}.
\eeq
A common way to obtain the solution of such a fix-point equation is to start 
with some initial guess, insert it in the right-hand side, to obtain an 
improved estimate on the left-hand side. If the function on the right-hand
side of (\ref{fix-point}) is well defined (i.e.\ the derivative is always 
positive) and contracting, then it follows from Banach's fix-point theorem that 
an iteration of this step will eventually converge to the MPLE. If $\cQ\equiv
\cL$, these conditions are satisfied and the resulting algorithm is known as 
Richardson (1972) - Lucy (1974) algorithm\footnote{
	The Richardson-Lucy algorithm can be generalized to maximize a more
	general functional, $\cA(f)$, subject to $\cN(f)=1$. Let $\cA(f_i)$
	the discretized form of $\cA(f)$, then the increment $\Delta f_i$ 
	for this algorithm can be obtained by requiring $\p\tcA/\p\ln f_i=0$ 
	(see equation \ref{eq-tcA-def} for the definition of $\tcA$), which 
	gives (\cite{luc94})
	\beq
		\Delta f_i = f_i \left({\p\cA\over\p f_i} -
			\sum_j f_j {\p\cA\over\p f_j}\right).
	\eeq
	However, for the algorithm to work one must have $\p\cA/\p f_i\ge0$ 
	everywhere, which places a strong restriction on the possible 
	applications; for instance, for $\cA=\cQ$ these conditions are
	generally not satisfied.}.
However, even when it works, the convergence will, in general, be no faster 
than linear.

Therefore, the maximization is more efficiently done by the conjugate-gradient 
algorithm (cf.\ Press \etal\ 1992). This technique too obtains the vector 
$\phi_\bsl$ of the MPLE, hereafter denoted by $\hphi_\bsl$, iteratively
(with ${\cal O}(L)$ operations per iteration). This time, however, the 
convergence will be quadratic, and the number of iterations theoretically
required is about $L$, so in total ${\cal O}(L^2)$ operations are required.
For $L\approx10^{5-6}$ this implies that each solution needs a considerable 
amount of computer time. 

\ifpreprint
  \begin{figure}
	\epsfxsize=8.6cm \epsfbox[20 155 325 710]{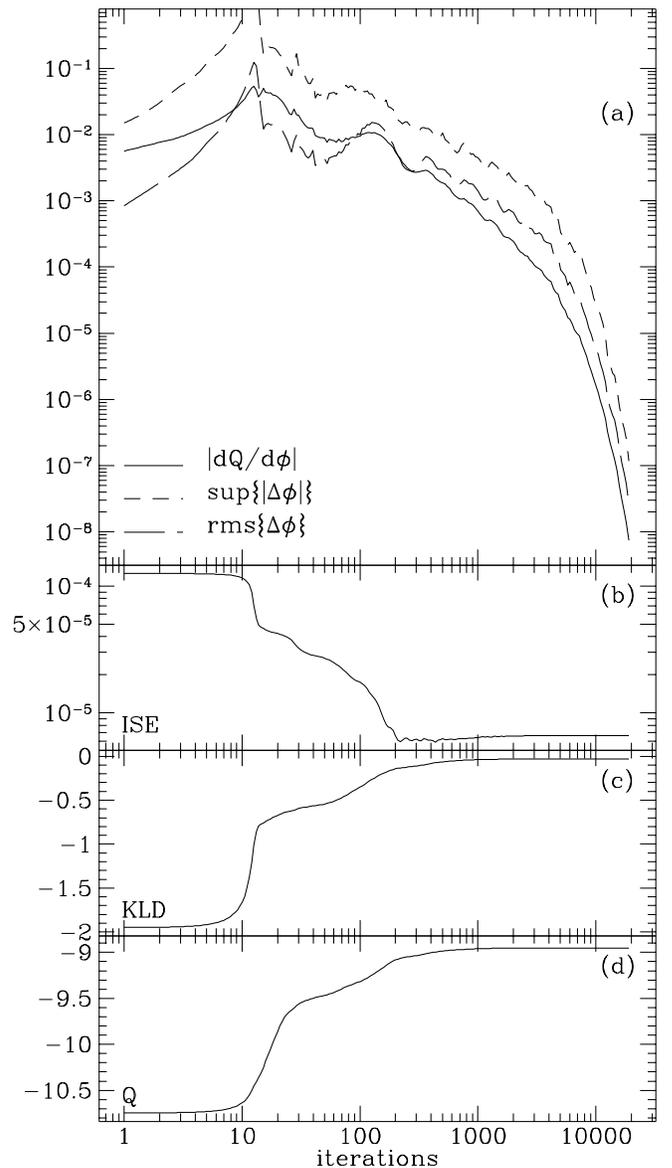}
	\caption[]{\footnotesize
	Conjugate-gradient maximization of $\tcQ$ for a run with $L=96\times
	80\times56$ and $N=1294$ data drawn at random from a model for 
	$f(\bvel)$. The panels show as a function of the number of iterations 
	completed: (a) gradient of \tcQ, supremum and {\sc rms} value of the 
	increment; (b) integrated square error (\ref{eq-ise}); (c) 
	Kullback-Leiber information-distance (\ref{eq-kld}); and (d) \cQ\
	itself.}
        \label{fig-conv}
  \end{figure}
\else
  \placefigure{fig-conv} 
\fi

Fortunately, experiments showed that already after much less iterations 
$\phi_\bsl$ is very close to $\hphi_\bsl$. For a run with $L=430\,080$ and 
$N=1294$ pseudo-data drawn from a model, the sum of three Gaussians, 
Figure~\ref{fig-conv} shows various quantities as function of the number of 
iterations completed. Already after $\sim10^4\approx0.02L$ iterations, the 
modulus of the gradient $\p\tcQ/\p\B{\phi}$ has decreased by four orders of 
magnitude (the iterations were actually stopped when it had decreased by 
$10^{-7}$); the same holds for both the supremum of the modulus and the {\sc 
rms}-value of the increment $\Delta\phi_\bsl$ at each iteration. For a rigorous
estimation of the error of the current $\phi_\bsl$ as compared to $\hphi_\bsl$,
one can extrapolate $\sup_\bsl\{|\Delta\phi_\bsl|\}$ into the future in order 
to estimate an upper limit for $\sup_\bsl\{|\phi_\bsl-\hphi_\bsl|\}$.

Panels (b) and (c) of Figure~\ref{fig-conv} show the run of the ISE and KLD 
between the current estimate for $f(\bvel)$ and the model underlying the 
pseudo-data. It is evident that after $\sim2000\approx0.005L$ iterations further
maximization of $\tcQ(f)$ does not change the quality of the current estimate, 
even though the convergence is still sub-quadratic (as judged from the run of 
$\sup_\bsl\{|\Delta\phi_\bsl|\}$). Clearly, it makes no sense to iterate far 
beyond this point. In practice, I iterate until $|\p\tcQ/\p\B{\phi}|$ falls 
below some threshold, e.g.\ $10^{-6}$, but perform at least a minimum number 
of iterations.

The number of operations can be considerably reduced by the multi-grid approach.
A solution obtained on a coarser grid is cheaper to obtain but contains already
a wealth of information. This can be exploited by starting on a really coarse 
grid, maximizing $\tcQ$ on this grid, transforming to a finer grid, and 
so on. I will use a sequence of four or five grids, each created from its 
predecessor by dividing the cells into eight daughter cells; $\B{\phi}$ is 
linearly interpolated to transform to the finer grid. Typically, on each of 
the grids 200 iterations are performed, except for the finest one, where the 
number might be higher. 

\ifpreprint
  \section{APPLICATION TO HIPPARCOS DATA}\label{sec-results} \nobreak
\else
  \section{Application to HIPPARCOS Data}\label{sec-results}
\fi
\subsection{The Subsamples}\label{sec-samples} \ifpreprint \nobreak \fi
Even though the \hip\ Catalogue contains many of the nearby stars,
its completeness varies with direction and stellar type 
and some care must be taken, in order to avoid kinematic biases, which are 
present in the full Catalogue and would render this study useless. In \paI, we 
have extracted from the \hip\ Catalogue a kinematically unbiased sample of 
11\,865 single main-sequence stars with parallaxes accurate to 10 percent or 
better, hereafter \saI. The kinematics inferred from this sample in \paI\ 
depend on color. For $\bv\lesssim0.61\magn$, the dispersion velocities increase 
for ever redder color bins, because, for main-sequence stars, the mean color 
correlates with the mean age, and scattering processes increase the random 
motions with time. Above $\bv\approx0.61\magn$ the dispersion is constant.
This transition is called Parenago's discontinuity and arises because at 
$\bv\ga0.61\magn$ subsamples of main-sequence stars have the same mean age.

To take this dependence of kinematics on color into account, I subdivide \saI\ 
into four color bins, labeled `B1' to `B4'. The first, B1, contains stars 
bluer than $\bv=0\magn$. These stars do not follow Str\"omberg's 
asymmetric-drift relation (the linear dependence of the mean $v_y$ on the 
velocity dispersion squared) defined by the other samples (Figure~4 of 
\paI). The fourth color bin, B4, consists of all stars of \saI\ red-ward of 
$\bv=0.6\magn$ and should be dominated by old-disk stars. The remaining two 
bins have intermediate colors. In addition to these bins of main-sequence 
stars, I consider a fifth set, labeled `GI', of stars, mainly giants, which 
are in the kinematically unbiased sample derived in \paI\ but excluded from 
\saI\ because they are off the main-sequence. Finally, the union of all these 
subsets, labeled `AL', is analyzed as a whole. So, in total there are six 
kinematically unbiased sets of stars, five of which do not overlap; more 
details are given in Table~\ref{tab-s}. 

The full set, AL, is magnitude limited and one should bear in mind that it
does not give a fair representation of the color and velocity distribution of 
typical stars in the solar neighborhood, but is heavily biased towards luminous 
stars. Subsets B1 to B4, however, being restricted to main-sequence stars of a
narrow color-range or beyond Parenago's discontinuity, are much less biased in
this sense. Finally, for the interpretation of the results it is useful to note
that the upper limit in \bv\ for the samples B1 to B3 of main-sequence stars 
places an upper limit on the stellar age; the last column of Table~\ref{tab-s} 
contains the values obtained from the stellar evolutionary models by Bressan 
\etal\ (1993) for a metallicity of $Z=0.02$.

\ifpreprint
  \begin{table}
  \small
  \caption[]{The subsamples analyzed \label{tab-s}}
  \smallskip

  \begin{tabular}{lrr@{\quad}r@{\quad}r@{\quad}c}
  name & \multicolumn{2}{c}{$(\bv)_{\rm min,max}$} & 
	$N_{\rm tot}$ & $N_{\rm in}$ & $\tau_{\rm max}$ \\[1ex] \hline
  B1& - &0.0&  524&  524&$4\times10^8$ \\
  B2&0.0&0.4& 3201& 3199&$2\times10^9$ \\
  B3&0.4&0.6& 4596& 4582&$8\times10^9$ \\
  B4&0.6& - & 3544& 3527&      -       \\
  GI& - & - & 2504& 2491&      -       \\
  AL& - & - &14369&14323&      -       \\[1ex] \hline	
  \end{tabular}
  \medskip

  \footnotesize
  Color limits (in mag), total number $N_{\rm tot}$ of stars, number $N_{\rm 
  used}$ of stars used, and maximum stellar age (in years), for the color
  bins B1 to B4 of main-sequence stars, the giant sample GI and the union of 
  these, AL. (see text).
  \end{table}

  \begin{figure*}
	\epsfxsize=\textwidth \epsfbox{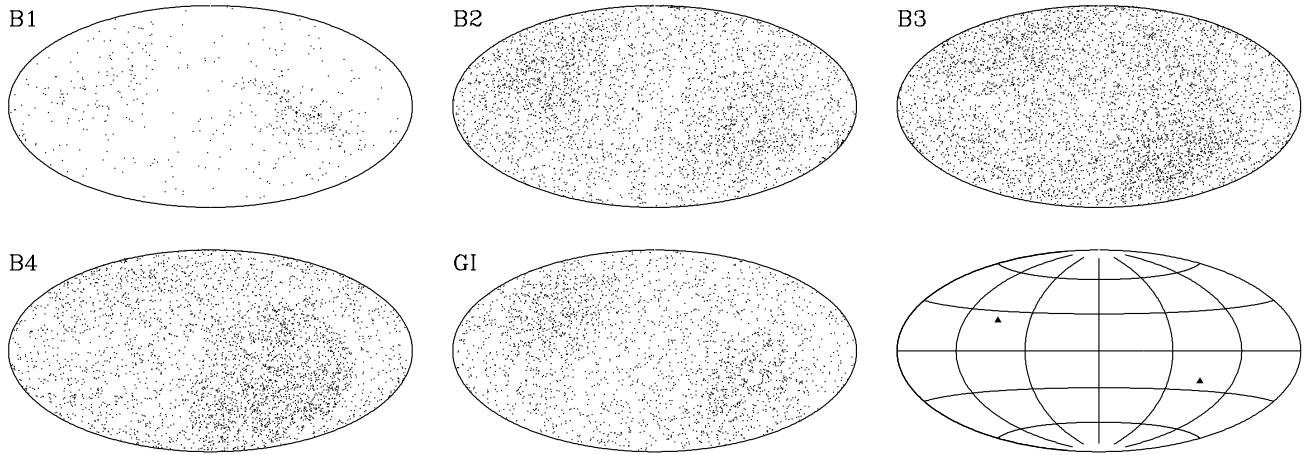}
	\caption[]{\footnotesize
	The data: distribution in galactic coordinates of the five distinct
	sets B1 to B4 and GI. The bottom right frame shows lines of constant 
	$\ell$ (from $180^\circ$ at the left to $-180^\circ$ at the right, 
	spacing of $60^\circ$) and $b$ (from $-90^\circ$ at the bottom to 
	$90^\circ$ at the top, spacing of $30^\circ$), the triangles mark the 
	poles of the ecliptic.}
	\label{fig-sky}
  \end{figure*}
\else 
  \placetable{tab-s}
  \placefigure{fig-sky} 
\fi

In Figure~\ref{fig-sky}, the positions on the sky of the stars in the five
distinct subsamples B1 to B4 and GI are displayed. Gould's Belt can be clearly
identified in B1, but the remaining subsamples do not show obvious clustering
(even the Hyades at $\ell\approx-175^\circ$, $b\approx-24^\circ$ are hardly 
detectable). However, the distributions are not uniform, there are more stars 
near the poles of the ecliptic. The reason is that at these poles the absolute
accuracies of the parallaxes measured by \hip\ are highest, so more stars 
survive our selection on relative distance errors. Another enhancement is 
clearly visible in the subsample B4 around the southern celestial pole 
(near the southern pole of the ecliptic). This is because, as outlined in \paI,
\saI\ contains stars from \hip\ proposal~018, which is restricted to 
declinations south of $-28^\circ$.

\subsection{The Distribution of Space Velocities}\label{sec-fofv}
For each of the six sets, $f(\bvel)$ was estimated as described in 
Section~\ref{sec-alg} with a finest grid of $96^3$ cells of size (velocities
are always given in \kms\ throughout this section) $3\times2.5\times1.8$ 
covering the box $[-154,134] \times[-160,80]\times[-93.4, 79.4]$. Stars whose 
$K(k|\bl)$ contributes to less than 96 cells are likely to originate from 
velocities outside this box and have been excluded from the analysis.

Because I have split the full sample into distinct subsamples, optimizing the
smoothing parameter (via the cross-validation technique of 
Section~\ref{sec-aopt}) for each subsample independently would erase information
that becomes significant only when more than one subsample is considered 
simultaneously: features that are consistent between two or more distinct 
subsamples are significant even if optimal smoothing would suppress these 
features in each of the subsamples. Therefore, I employed the cross-validation 
technique only to optimize the smoothing for the full sample (AL). The resulting
values $\alpha=3.42 \times10^{-10}$ and $\tilde{\B{\sigma}}=(30.6,21.3,14.6)$ 
(obtained as the velocity dispersion computed via the method of \paI) minimize 
the MISE and MKLD at $(2.13\pm0.09)\times10^{-6}$ and $0.0071\pm0.0003$, 
respectively. For the subsets, except B1, I fixed $\alpha$ and $\tilde{\B{\sigma
}}$ at these values (which means that $\alpha$ is up to about four times smaller
than the optimal value for each subset). Subset B1 has much fewer stars in it 
than the other subsets and taking these values for the smoothing parameters 
resulted in a greatly under-smoothed estimate; I therefore used $\alpha=1.78
\times10^{-9}$ for subset B1.

As a consistency check, I evaluated the first and second velocity moments from 
the inferred $f(\bvel)$; they agreed well with the moments estimated 
via the method of \paI. To test whether contamination with known clusters or 
associations is a serious problem, I applied the algorithm to subsamples with 
the region on the sky around the Hyades cluster and the region dominated by 
Gould's Belt (for B1) being excluded. The results were similar to that obtained
without excluding these stars.

\ifpreprint
  \begin{figure*}
	\centerline{\epsfxsize=8.45cm\epsfbox[25 32 310 237]{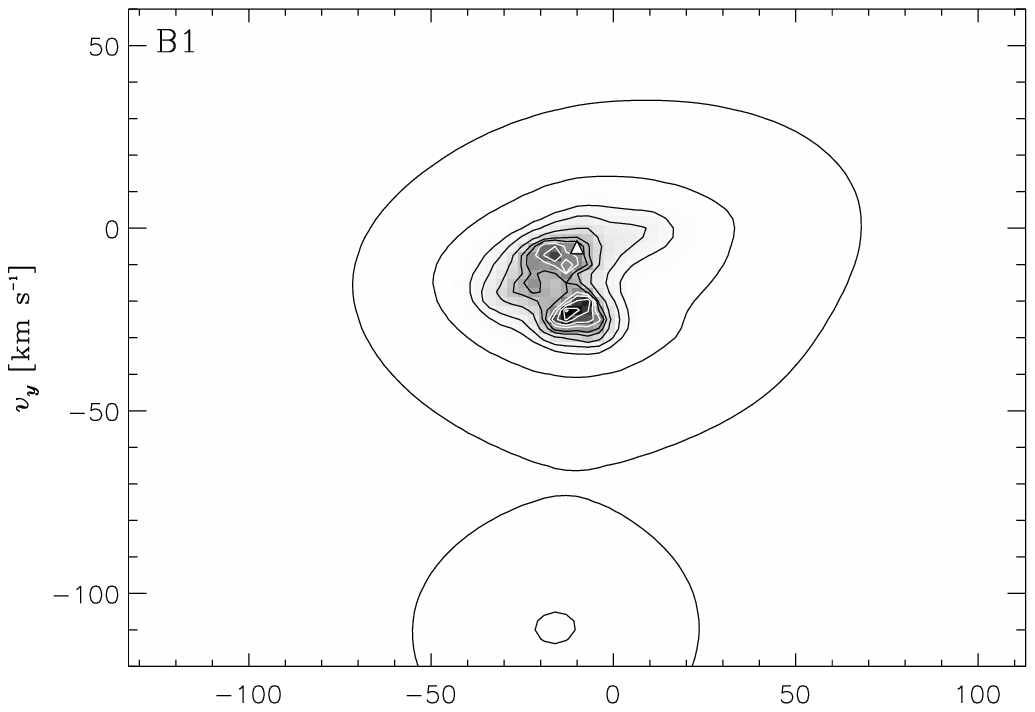}
		    \epsfxsize=8.45cm\epsfbox[25 32 310 237]{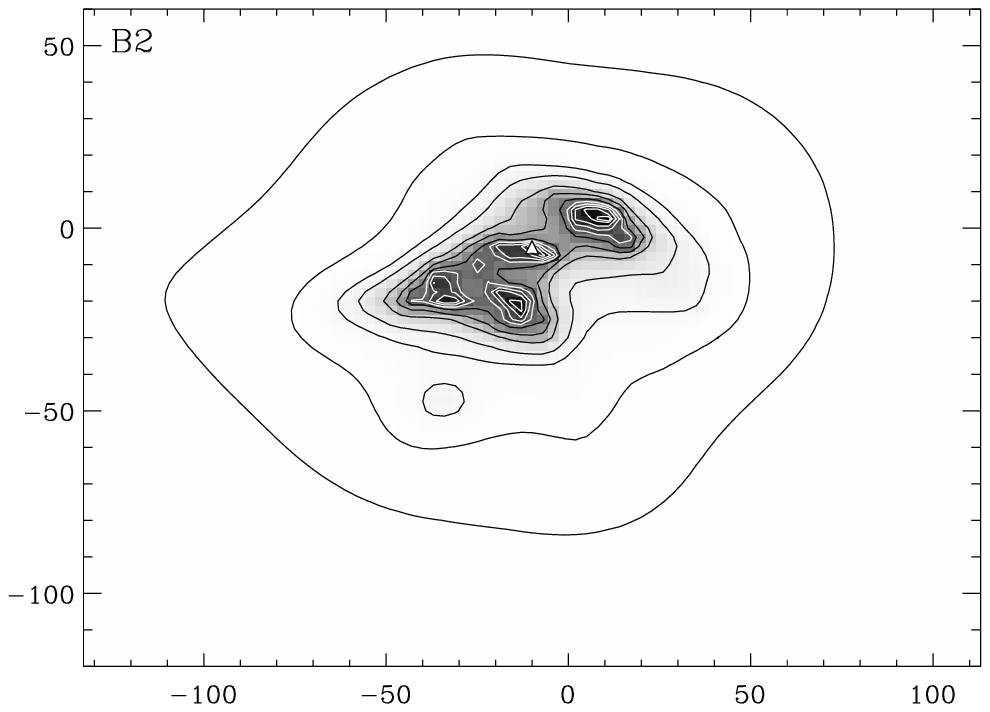}}
	\centerline{\epsfxsize=8.45cm\epsfbox[25 32 310 237]{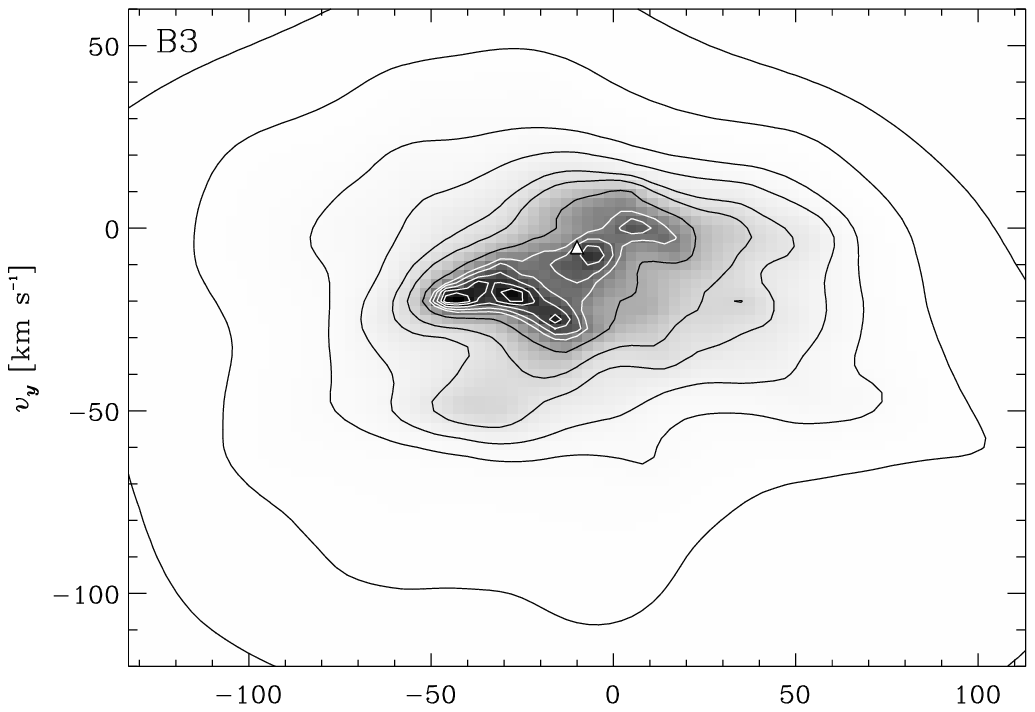}
		    \epsfxsize=8.45cm\epsfbox[25 32 310 237]{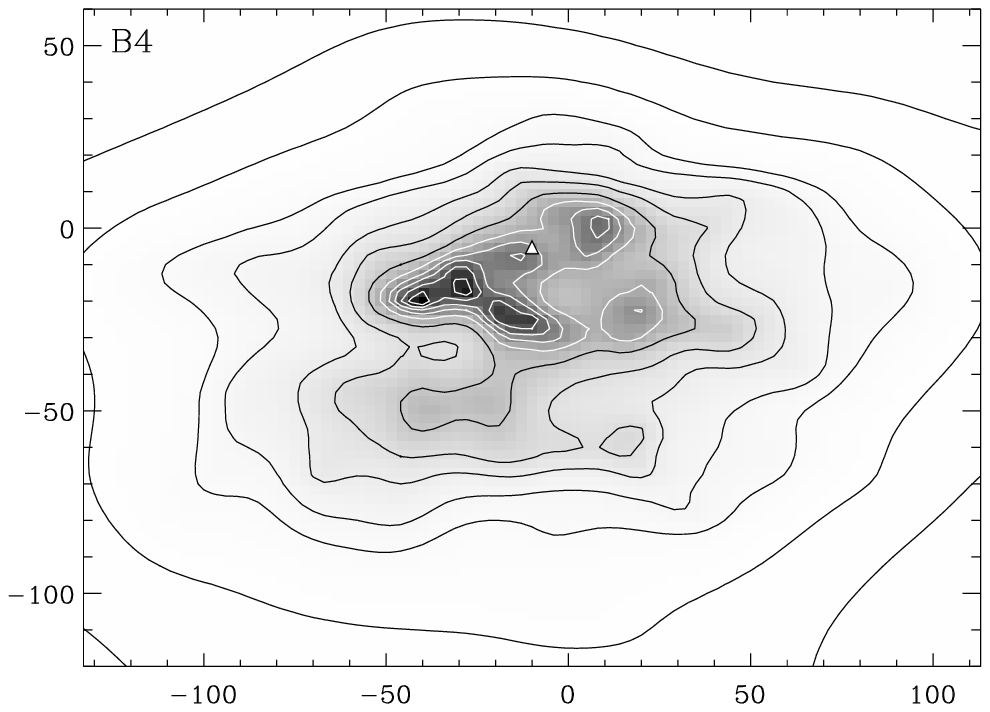}}
	\centerline{\epsfxsize=8.45cm\epsfbox[25 12 310 237]{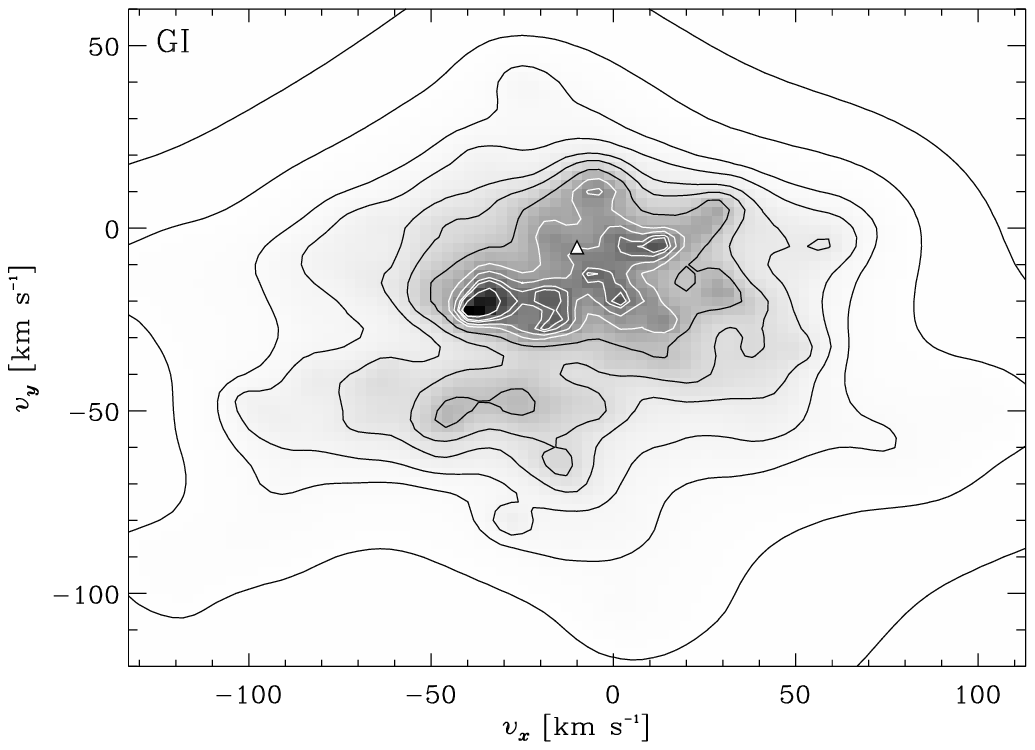}
		    \epsfxsize=8.45cm\epsfbox[25 12 310 237]{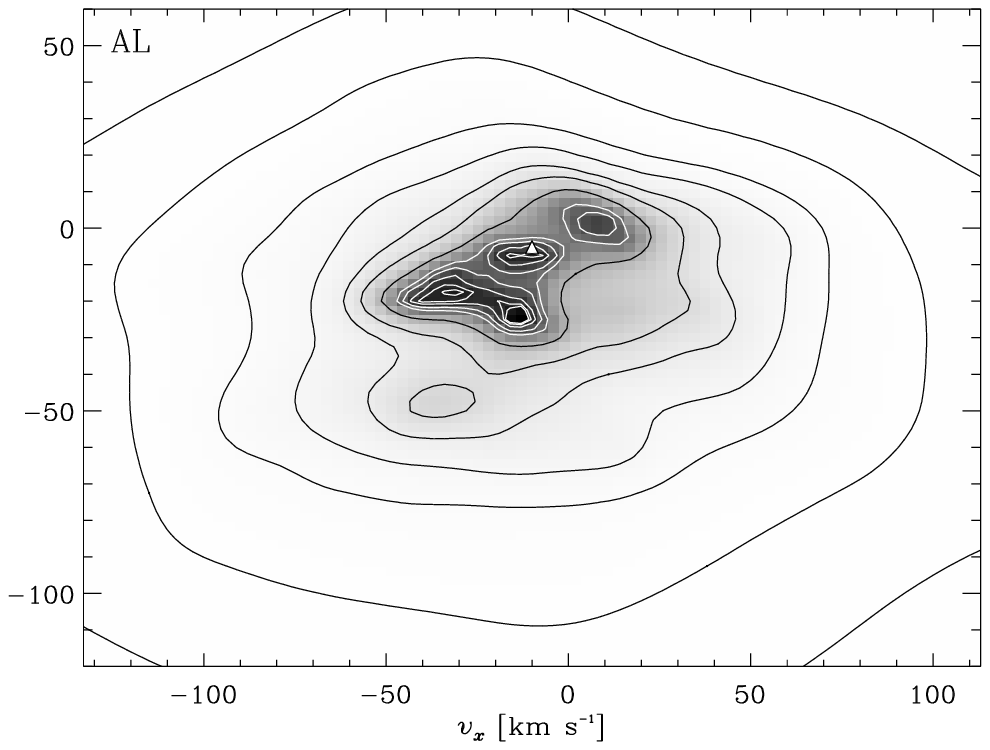}}
	\caption[]{\footnotesize
	The distributions in $v_x$ (towards the Galactic center) and $v_y$
	(in direction of Galactic rotation): projection of $f(\bvel)$ obtained
	as MPLE for the sets listed in Table~\ref{tab-s}. Gray scales are linear
	and the contours contain, from inside outwards, 2, 6, 12, 21, 33, 50, 
	68, 80, 90, 95, 99, and 99.9 percent of all stars, i.e.\ half the 
	stars are within the innermost dark contour. The origin is at the 
	solar velocity, while the velocity derived for the LSR in \paI\ is 
	indicated by a triangle. Note that the smoothing is optimal for the 
	full sample (AL) only, while the results for the subsets are 
	under-smoothed. However, since the subsets are distinct, any feature 
	common to more than one of them is likely to be real.}
	\label{fig-fuv}
  \end{figure*}
  \begin{figure*}
	\centerline{\epsfxsize=8.45cm\epsfbox[25 32 310 194]{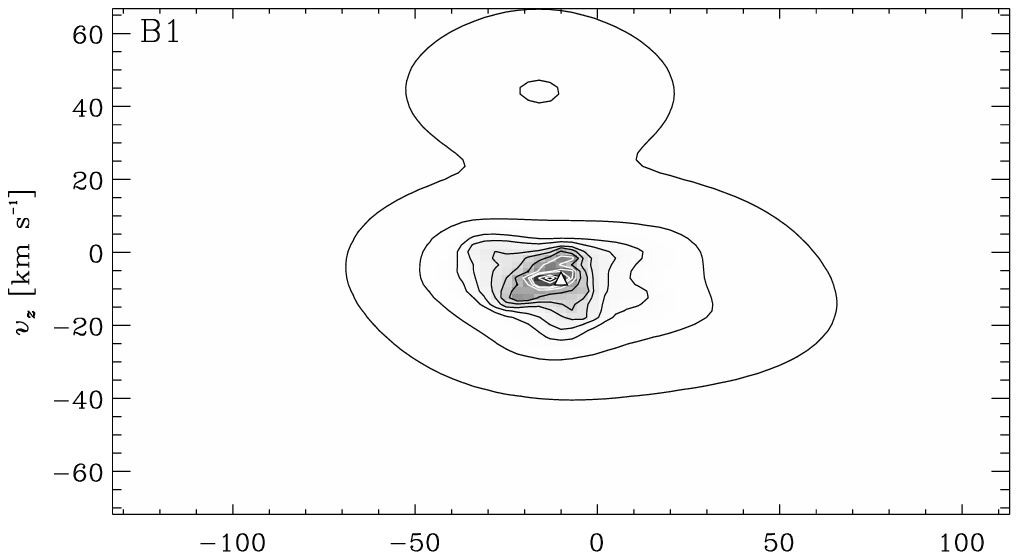}
		    \epsfxsize=8.45cm\epsfbox[25 32 310 194]{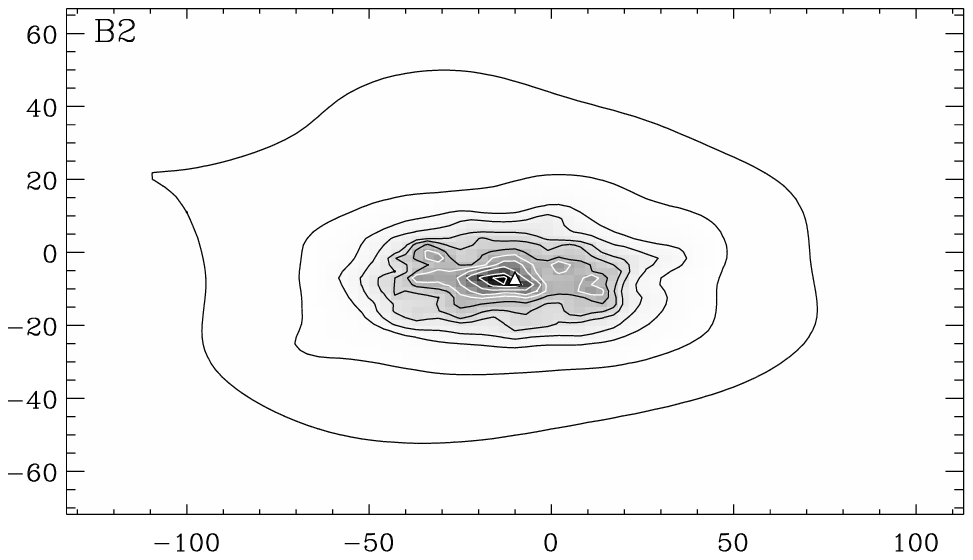}}
	\centerline{\epsfxsize=8.45cm\epsfbox[25 32 310 194]{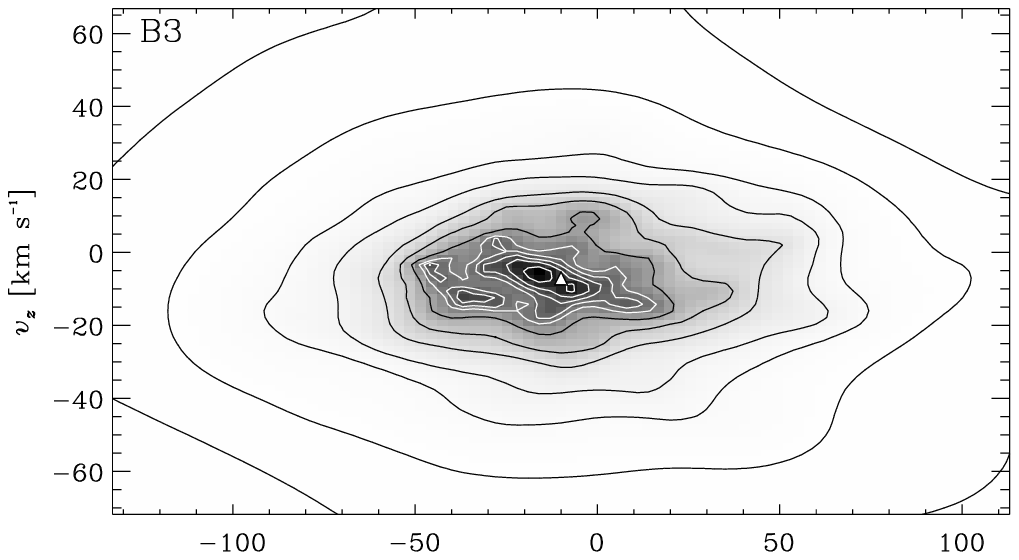}
		    \epsfxsize=8.45cm\epsfbox[25 32 310 194]{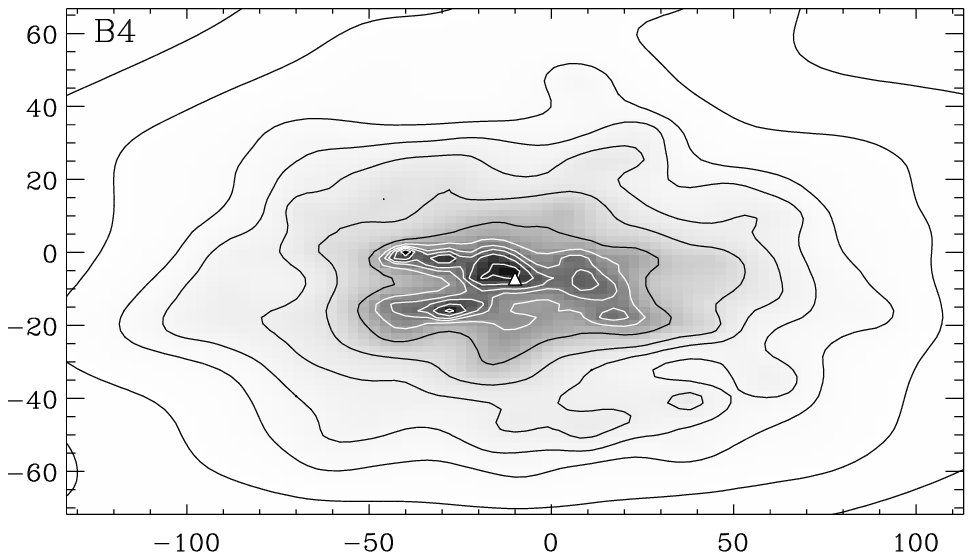}}
	\centerline{\epsfxsize=8.45cm\epsfbox[25 12 310 194]{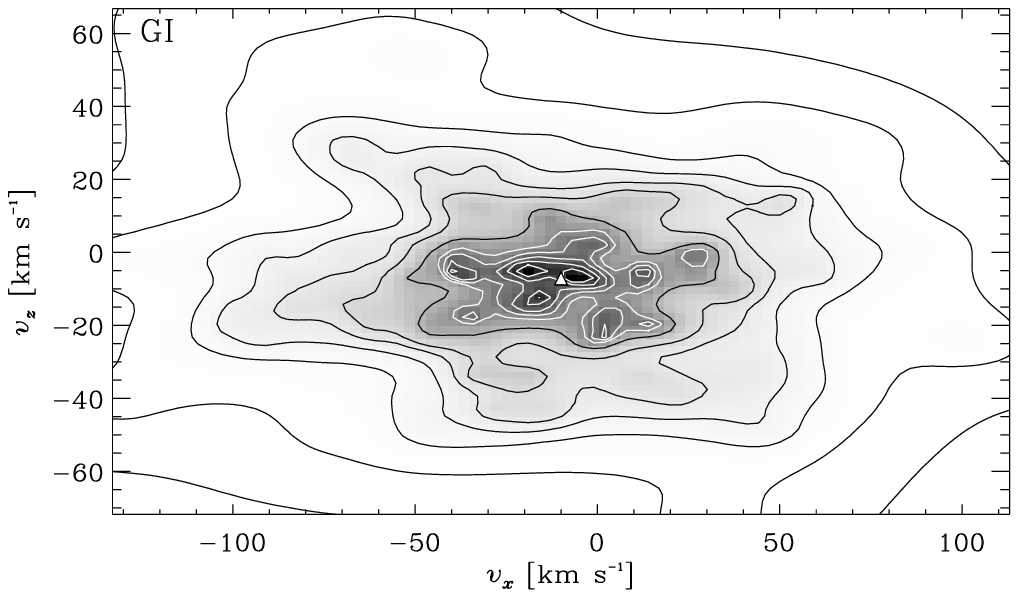}
		    \epsfxsize=8.45cm\epsfbox[25 12 310 194]{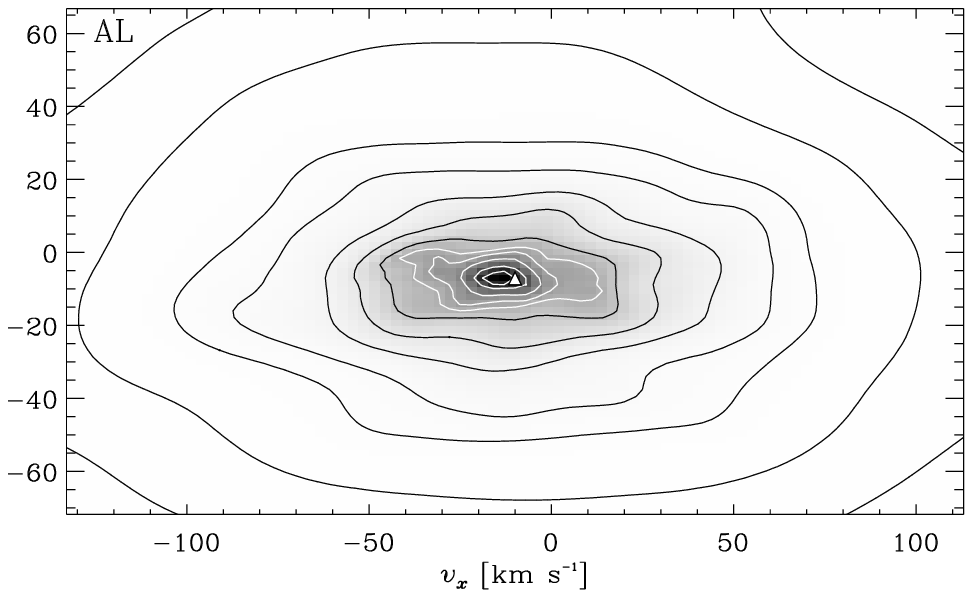}}
	\caption[]{\footnotesize
	The distributions in $v_x$ (towards the Galactic center) and $v_z$
	(towards the NGP). See Fig.~\ref{fig-fuv} for the contour levels, etc.}
	\label{fig-fuw}
  \end{figure*}
  \begin{figure*}
	\centerline{\epsfxsize=6.19cm\epsfbox[25 32 310 254]{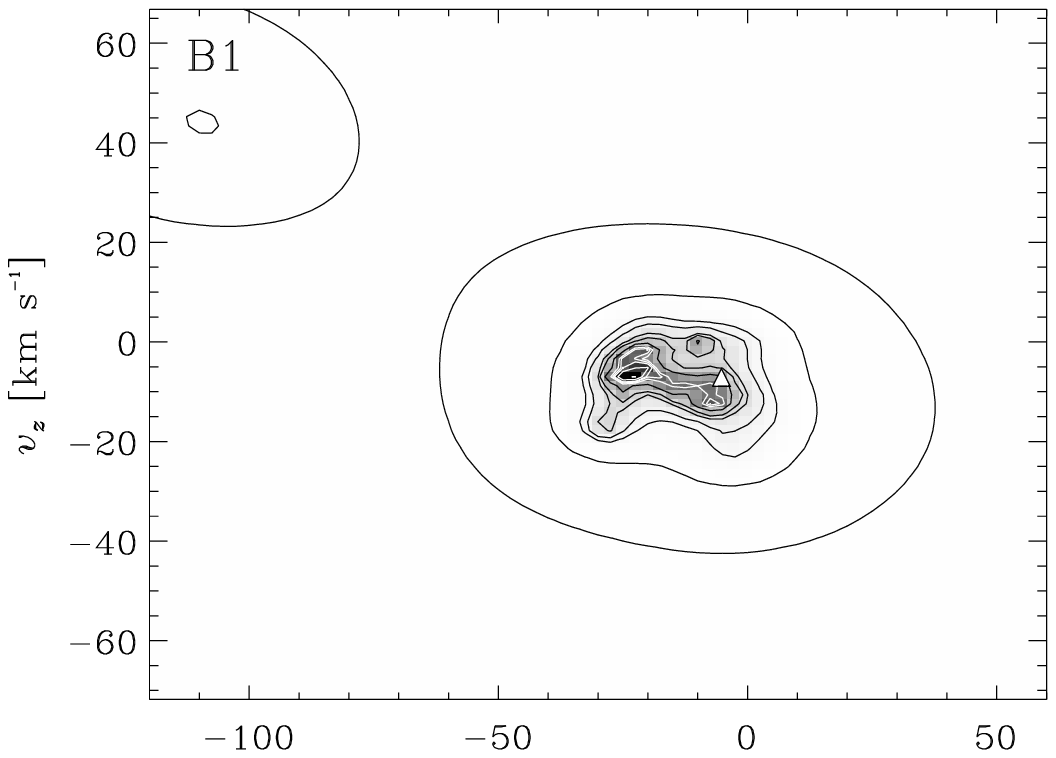}
		    \epsfxsize=6.19cm\epsfbox[25 32 310 254]{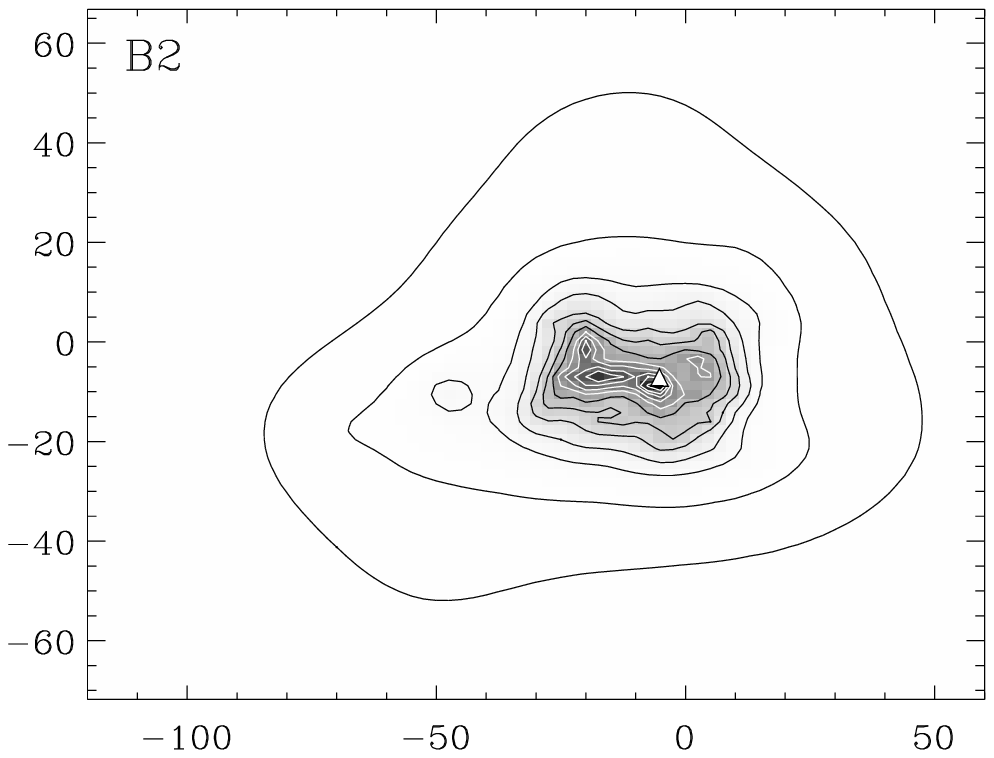}}
	\centerline{\epsfxsize=6.19cm\epsfbox[25 32 310 254]{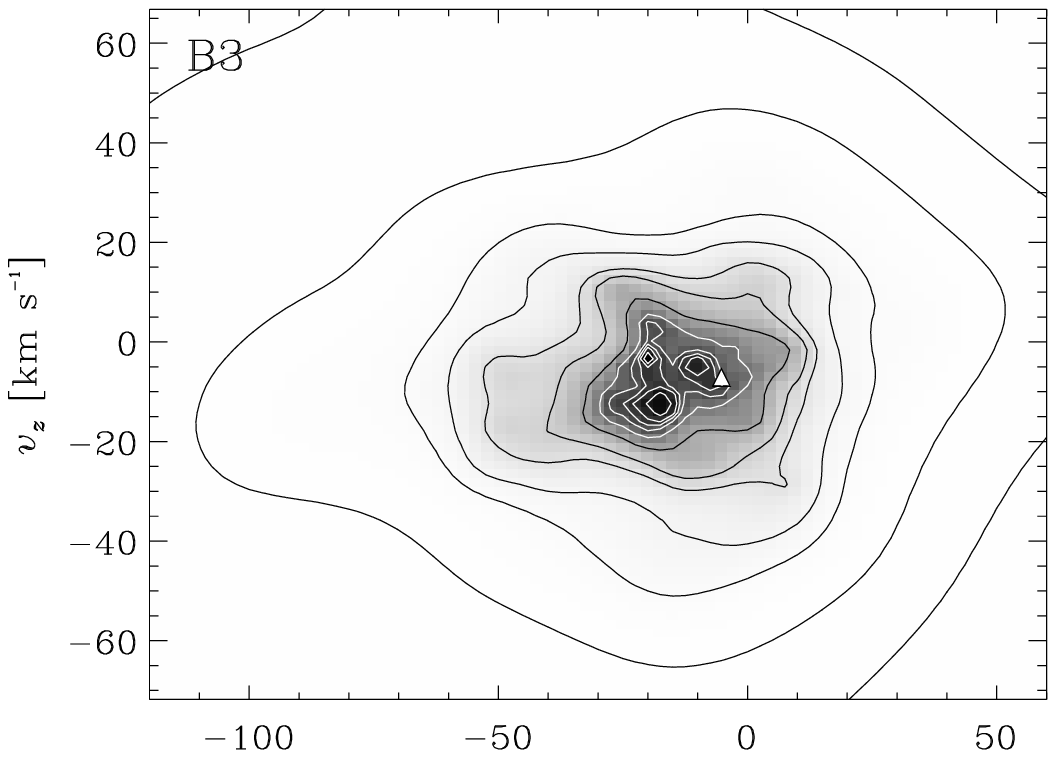}
		    \epsfxsize=6.19cm\epsfbox[25 32 310 254]{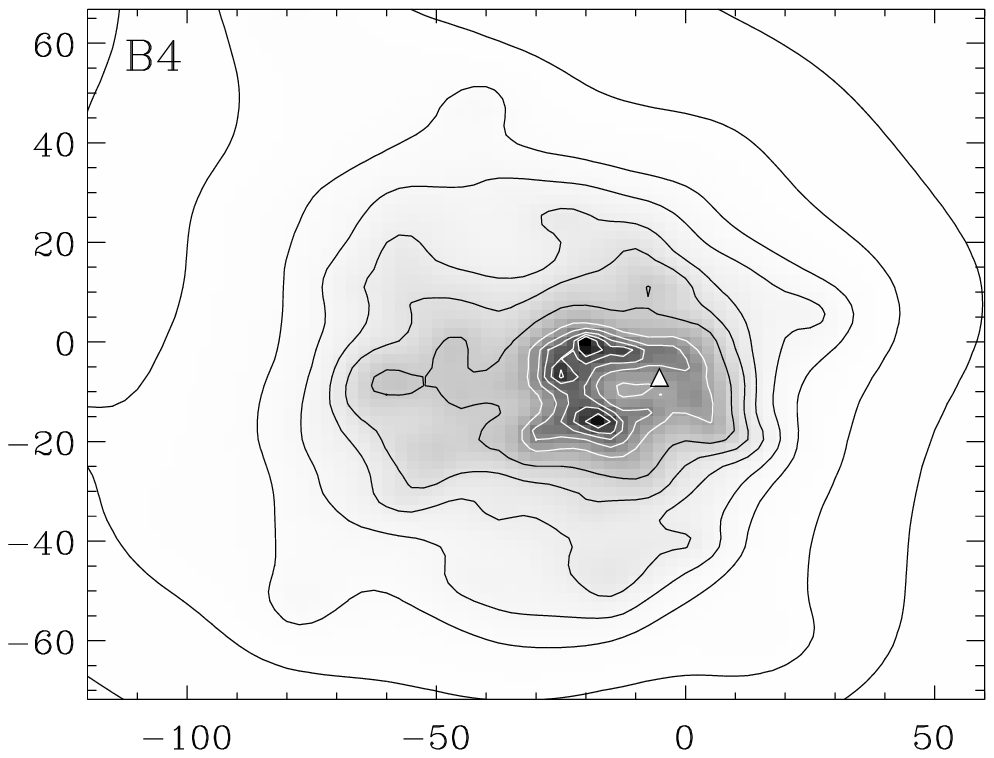}}
	\centerline{\epsfxsize=6.19cm\epsfbox[25 12 310 254]{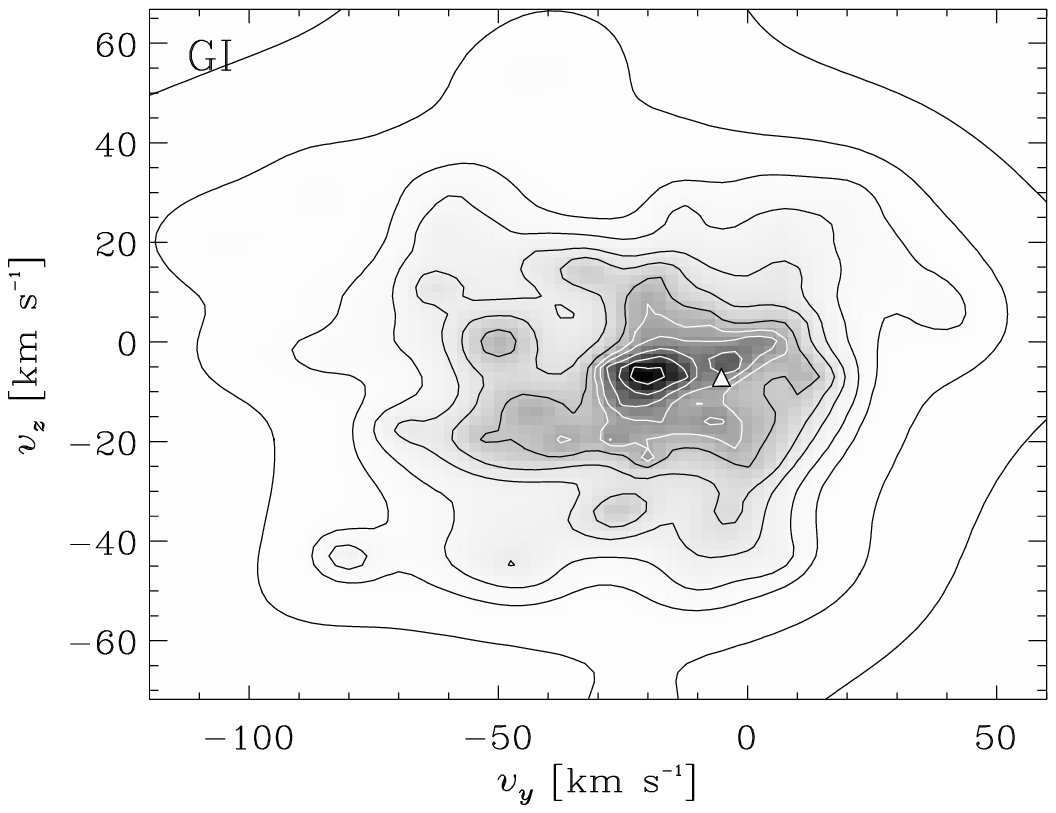}
		    \epsfxsize=6.19cm\epsfbox[25 12 310 254]{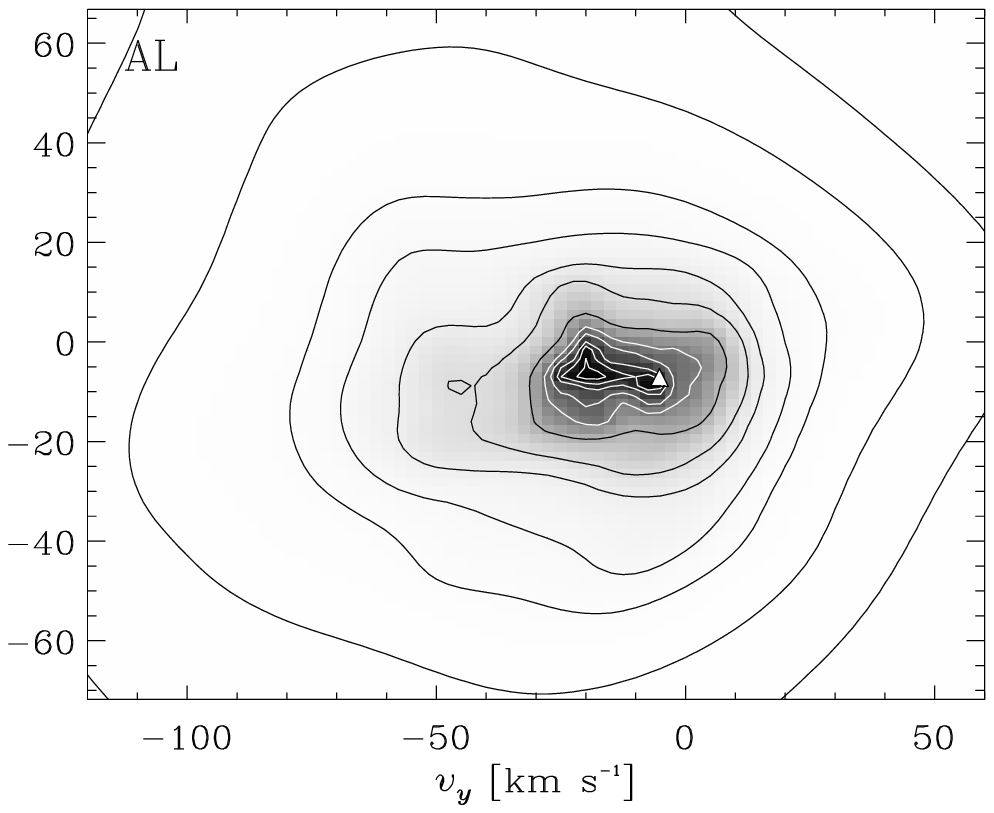}}
	\caption[]{\footnotesize
	The distributions in $v_y$ (in direction of Galactic rotation) and $v_z$
	(towards the NGP). See Fig.~\ref{fig-fuv} for the contour levels, etc.}
	\label{fig-fvw}
  \end{figure*}
\else
  \placefigure{fig-fuv}
  \placefigure{fig-fuw}
  \placefigure{fig-fvw}
\fi

For the six sets of Table~\ref{tab-s}, Figures~\ref{fig-fuv}, \ref{fig-fuw},
and \ref{fig-fvw} show the projections of the estimated $f(\bvel)$ onto the
$v_xv_y$, $v_xv_z$, and $v_yv_z$ plane, respectively. 

\ifpreprint
  \begin{table}
  \small
  \tabcolsep.5em
  \def\y{\multicolumn{1}{c}{$\bullet$}}
  \def\q{\multicolumn{1}{c}{$\circ$}}
  \caption[]{Features in the inferred velocity distributions\label{tab-st}}
  \smallskip

  \begin{tabular}{r@{\quad}r@{\hspace{0.5em}}r@{\hspace{0.5em}}r@{\quad}%
        	  c@{}c@{}c@{}c@{}c@{}c@{\quad}l}
  No. & $v_x$ & $v_y$ & $v_z$ & B1&B2&B3&B4&GI & & moving group \\[1ex] \hline
   1&$-12$&$ -22$&$ -7$&\y&\y&\y&\y&\q&&Pleiades          \\
   2&$-40$&$ -20$&$  0$&  &\y&\y&\y&\y&&Hyades            \\
   3&$  9$&$   3$&$   $&\y&\y&\y&\y&\y&&Sirius \& UMa     \\
   4&$-10$&$  -5$&$ -8$&\y&\y&\y&\y&  &&Coma Berenices    \\
   5&$-25$&$ -10$&$-15$&  &\y&\y&\y&\q&&NGC\,1901         \\
   6&$ 20$&$ -20$&$   $&  &  &  &\y&\y&&                  \\
   7&$ 15$&$ -60$&$   $&  &  &  &\y&\q&&HR1614            \\
   8&$-40$&$ -50$&$   $&  &\y&\y&\y&\y&&                  \\
   9&$-25$&$ -50$&$   $&  &  &\q&\y&\y&&                  \\
  10&$  0$&$-110$&$   $&\y&  &\q&\q&\q&&Arcturus          \\
  11&$-70$&$ -10$&$   $&  &  &\q&\q&\q&&                  \\
  12&$-70$&$ -50$&$   $&  &  &\q&\q&\y&&                  \\
  13&$ 50$&$   0$&$   $&  &  &\q&\y&\q&&                  \\
  14&$ 50$&$ -25$&$   $&  &  &\q&\y&\q&&                  \\[1ex] \hline
  \end{tabular}
  \medskip

  \footnotesize
  Velocities are approximate ($v_x$ is $-U$ in Eggen's papers); if $v_z$ is 
  missing it cannot be determined. $\bullet$ and $\circ$ denote, respectively, 
  clear or vague visibility in the corresponding subsample. The list of 
  associated moving groups is presumably incomplete.
  \end{table}
\else
  \placetable{tab-st}
\fi

The most conspicuous characteristic of the inferred velocity distributions
are several strong maxima and many minor wiggles most apparent in the planar 
motions (Fig.~\ref{fig-fuv}). These features are less clear but still 
significant in the sample AL, for which optimal smoothing has been used.
Almost all of these major and minor maxima and even some of the wiggles can be 
identified with one of Eggen's (1965, 1995, 1996) moving groups. 
Table~\ref{tab-st} lists the positions of 14 such features in 
Figures~\ref{fig-fuv} to \ref{fig-fvw}, together with the name of an associated 
moving group, where one could be found in the literature. (Some of the 
associations might well be wrong, for instance, Eggen gives an age of 10 Gyr 
for the Arcturus group, much more than any star possibly in B1.) The most 
prominent of these features (i.e.\ the first 4-5 in that table) have also been 
found by other groups from analysis of \hip\ data, e.g.\ Figueras \etal\ (1997)
using \hip\ data in conjunction with radial velocities for young B5-F5 stars, 
and Chereul \etal\ (1997) using a convergent-point method (see footnote 1) for 
3000 A stars\footnote{The latter authors even 
	claim substructure on a scale of a few $\mkms$, which I must call into 
	question in the light of tests I have made: on small scales noise 
	amplification inevitable creates spurious structures -- a phenomenon 
	characteristic for inverse problems.}. 
The distribution of these maxima is skewed in the $v_xv_y$ plane (but not
in the other two principal planes), which in turn is responsible for 
the tilt in the velocity ellipsoid (vertex deviation) derived in \paI.

As one moves through the sequence B1 to B4, i.e.\ considers ever redder
and hence on average older main-sequence stars, the following can be noticed.
\begin{itemize}
\item[(1)] The relative number of stars in the maxima diminishes. While roughly
	half of the stars in subsamples B1 and B2 are associated with one of 
	the features, in B4 over $\sim3/4$ cent are in a smooth background 
	distribution.
\item[(2)] The number of features increases. This means that some of the moving
	groups are old and enter the red subsamples only, whereas young 
	clusters have stars of all colors.
\item[(3)] Moving groups that are present only in the red subsamples
	are predominantly at large negative azimuthal velocities. Thus there
	is a correlation in the sense that the older a moving group is, the 
	smaller is its (local) rotational velocity around the Galaxy. Similarly,
	most features at small or positive $v_y$ decrease in amplitude as one 
	moves to redder samples. 
\item[(4)] The width of the distribution of moving groups clearly increases in 
	$(v_x,v_y)$ but much less so in $v_z$. Together with (3) this means 
	that the distribution of moving groups obeys an asymmetric drift 
	relation.
\item[(5)] The width of individual maxima increases (at least for the four
	most prominent ones).
\item[(6)] From B2 to B4, the extent of the outermost two contours (containing 
	99 and 99.9 per cent of the stars) increases by a factor of about two 
	in $v_x$ and $v_y$ and about three in $v_z$, while the center is 
	shifted to more negative $v_y$, reflecting the asymmetric drift.
\item[(7)] From B2 to B4, the axis ratio of the outer contours changes from
	about 1:0.6:0.35 to 1:0.6:0.5. A similar change, however, is also 
	visible in each individual subsample when one moves from inside (small 
	$\bvel$) outwards.
\end{itemize}
With respect to all these points, subsample GI of non-main-sequence stars 
most closely resembles sample B4 of main-sequence stars redder than $\bv=
0.6\magn$.

It should be noted that, in contrast to the projections onto the $v_xv_y$ plane,
most features apparent in $v_z$ seem not be real (they do not occur in more than
one distinct subsample). In fact, they all disappear when optimal smoothing is
applied (not shown) as is the case for sample AL, with the remarkable exception
of the double peaked $f(v_z)$ at $(v_x,v_y)\approx(-30,-20)$ in B4.

\ifpreprint
  \section[]{MOVING GROUPS AND \\ STELLAR STREAMS}
\else
  \section{Moving Groups and Stellar Streams}
\fi
About half of the stars in the blue color bins B1 and B2 (with $\bv\le0.4\magn$)
are associated with moving groups, but the same holds for a considerable
fraction (up to $\sim25$ per cent) of main-sequence stars in the red color bins 
B3 and B4, even red-ward of Parenago's discontinuity at $\bv\approx0.61\magn$ 
(\paI), where no disk star has yet left the main sequence. This means that many
of the moving groups, in particular the less prominent ones, must be made of 
stars older than 2-8 Gyr, the ages of the oldest stars possibly in samples B2 
and B3, respectively.

\ifpreprint
  \subsection[]{The Shape of Moving Groups\\ in Velocity Space}
\else
  \subsection{The Shape of Moving Groups in Velocity Space}
\fi
According to the standard picture of these phenomena, stars form in clusters
each of which is initially confined to a small volume in phase-space. 
Over time these dissolve, such that the old stellar population obeys a 
smooth distribution both in configuration and velocity space. While the 
mechanisms driving the evolution and dissolution of the initially bound cluster
have been studied in some detail (\cite{ter87}), little is known about the 
subsequent evolution. Once the cluster becomes unbound, its stars will have
slightly different orbits and hence different frequencies, such that they
will phase-mix. A cluster with initial velocity dispersion $\sigma_0$ will be
dispersed over all azimuths in a time of $2\pi R/\sigma_0$ ($\sim5\,$Gyr for
$R=R_0$ and $\sigma_0\approx10\kms$). Since, at any fixed azimuth one 
expects only stars with identical azimuthal frequency, i.e.\ identical angular 
momentum, such a stellar stream is expected to manifest itself as a clump
in $f(\bvel)$ that is narrow in $v_y$ but has size $\gtrsim\sigma_0$ in 
$(v_x,v_z)$ (\cite{wol61})%
\footnote{Of course, in reality the stars of my sample are not exactly at the 
	solar position but have a {\sc rms} deviation of $d\approx40\parc$
	in each direction. This leads to a blurring of the clumps, since stars 
	on identical orbits will have different velocities throughout the 
	sampling volume. For an axisymmetric galaxy with flat rotation curve, 
	the size of this blurring can be estimated from conservation of energy, 
	angular momentum, and vertical energy to be 
	\ben
		\sigma_x &\approx& v_0\sqrt{d/R_0} \\
		\sigma_y &\approx& v_0\,d/R_0 \\	\label{eq-sigma_z}
		\sigma_z &\approx& \sqrt{\Phi(d) - \Phi(0)}
	\een
	where $v_0$ is the circular speed and $\Phi(z)$ vertical Galactic 
	potential. With $R_0=8\kpc$, $v_0=200\kms$, and using model 2 from 
	Dehnen \& Binney (1998a) for the potential of the Milky Way in 
	(\ref{eq-sigma_z}) this gives $\B{\sigma} \approx (14,\,1,\,3) \kms$. 
	Adding this quadratically to the effect of phase-mixing amplifies the 
	expected elongation of the clumps in $\bvel$-space.}.
Although measurement uncertainties and noise tend to sphericalize these clumps,
it is somewhat surprising, that only a few of the features in 
Figure~\ref{fig-fuv} are elongated in this sense. However, one also expects 
that scattering with perturbers, like giant molecular clouds or spiral arms, 
will randomly change the stellar orbits and hence spread the stream deleting 
the correlations in velocities.
On the other hand, as has long been known, the moving groups, except the very 
young ones, are almost indistinguishable in $v_z$. Presumably, this can be 
attributed to phase-mixing which is more efficient than for the horizontal 
motions, because, for any realistic disk-profile, the vertical frequency is a 
strong function of vertical energy, and the dynamical time is shorter by about
a factor of two.

\subsection{Moving Groups on Eccentric Orbits}
Thus, it appears natural that there is much more structure in $(v_x,v_y)$ than 
there is in $v_z$, however, the character of this structure is very interesting.
In particular, the fact that the distribution of moving groups obeys an 
asymmetric drift relation (points 3 \& 4 in Section~\ref{sec-fofv}), similar
to the smooth background: the older groups are more wide spread in $\bvel$ and 
lag w.r.t.\ the LSR, i.e.\ they are on non-circular orbits. For stars in the 
smooth background these relations arise because scattering processes increase 
the velocity dispersion with age, and because much more stars visit us from 
inside the solar circle than from outside. The latter certainly also holds for 
the moving groups and causes the lag. However, scattering processes would 
destroy and disperse an existing moving group rather than shifting it to another
orbit. The question therefore arises, how did the old moving groups get on their
eccentric orbits? 

Even tough one could imaging a scenario where the moving groups have been born 
onto these orbits -- for instance, by star formation in the molecular gas ring 
at $\sim4\kpc$, where the Galactic bar significantly distorts the axisymmetry 
-- this seems unlikely in view of the facts that phase-mixing should have 
spread these moving groups in $v_x$ and that the apparently young moving 
groups are indeed on near-circular orbits. Alternatively, the moving
groups could have been transformed to more eccentric orbits. This would
naturally account for the age-dependence of the eccentricity. Since a moving
group is a rather fragile object, any non-axisymmetric force field responsible 
for such a process must be smooth, both in space (long wavelength) and time 
(low frequency). While a triaxial halo or the Magellanic clouds might play a 
role in such an process, the most obvious candidate is the Galactic bar.
Actually, almost all the apparently older moving groups have negative $v_y$, 
and thus low $L_z$, they are near the apocenters of their orbits, which probe
a large range of radii and come near to the bar.

A plausible mechanism that is able to shift moving groups to eccentric 
orbits has been described by Sridhar \& Touma (1996). Applied to our problem
it works as follows. If the otherwise axisymmetric Galaxy is perturbed by some
non-axisymmetric component, its phase-space structure will change, the 
degeneracy will be broken, and resonant islands will occur, regions in 
phase-space surrounding resonant orbits. The orbits in these islands are 
near-resonant and oscillate around their resonant parent orbit. When the Galaxy 
slowly evolves, for instance, when the Galactic bar slows down and strengthens,
these resonant islands sweep through phase-space, and stars trapped in them 
will be shifted to different orbits -- an effect also known from the dynamics 
of the solar system. 

Actually, there is a clear pointer to the involvement of resonances in the 
motion of nearby stars: the recent re-determination of Oort's constants from 
\hip\ Cepheids by Feast \& Whitelock (1997) gives (in \kmskpc) $\Omega= 
A-B=27.2\pm0.9$, $\kappa=2\sqrt{B^2-AB}=36.7\pm1.6$, and $\Omega\,:\,\kappa=
2.97\pm0.07\,:\,4$. That is very precisely a 3:4 resonance between the azimuthal
and radial orbital frequency, a ratio which appears naturally in a slightly 
declining circular-speed curve, e.g.\ of the form $v_c\propto R^{-1/9}$. 3:4 
resonant orbits complete four radial oscillations in the same time they rotate
three times around the Galactic center; during this time they visit any given 
azimuth with three different velocities. In this respect it is intriguing that 
the stars in B1, which are at most two orbital rotation periods old, show just
two distinct peaks in their velocity distribution.

Currently, the only model that aims to explain the structure in $(v_x,v_y)$
does in fact invoke a resonance, although, not the one mentioned above.
This model is due to Kalnajs (1991), who starts from the observation of just
two major maxima (at the Sirius and Hyades peaks), and locates the Sun at the
outer Lindblad resonance (OLR) of the Galactic bar. The closed orbits inside
and outside the OLR are elongated, respectively, perpendicular and parallel to
the bar. At some azimuths these differently orientated orbits cross and 
naturally create a double-peaked distribution in $(v_x,v_y)$. This model gives 
pattern speed and projection angle of the Galactic bar in rough agreement with
other methods, but does not account for the asymmetric drift relation amongst
moving groups.

\ifpreprint
  \section{EVIDENCE FOR THE STELLAR WARP}
\else
  \section{Evidence for the Stellar Warp}
\fi
In Figure~\ref{fig-fvw}, in all samples except B1 and best visible in GI
and AL, the few outermost contours are somewhat skewed in the sense that at
positive $v_y$ more stars are moving upwards w.r.t.\ the LSR than downwards.
This is nicely illustrated in Figure~\ref{fig-vz}, where the mean vertical 
motion due to $f(\bvel)$ inferred from the full sample is plotted versus $v_y$. 
While $\bar{v}_z$ is roughly constant at the LSR value of $-7\kms$ for 
$v_y\la10\kms$ (the bends are presumably due to the moving groups), it 
clearly curves upwards for $v_y\ga10\kms$ (this is not an ordinary tilt, which 
would produce a straight line in this diagram).

\ifpreprint
  \begin{figure}
	\epsfxsize=8.6cm \epsfbox[20 320 600 720]{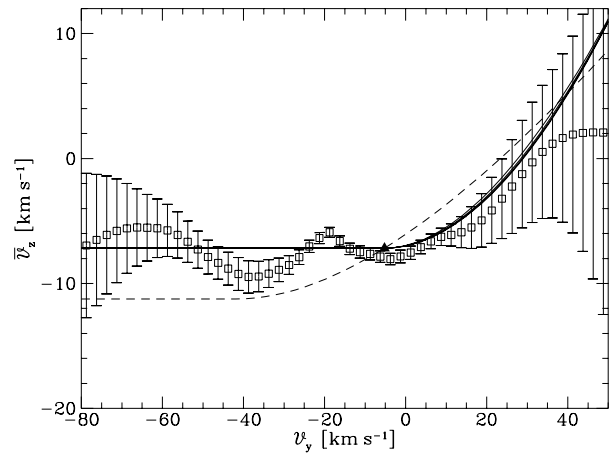}
	\caption[]{\footnotesize
	The mean vertical motion, $\bar{v}_z$, measured from sample AL as
	a function of of $v_y$. The errors bars show 1 $\sigma$ errors due to 
	Poisson noise; however, adjacent points are not independent but coupled
	by the smoothing. The triangle indicates the velocity of the LSR, and 
	the lines are derived from models for the Galactic warp, see text.}
	\label{fig-vz}
  \end{figure}
\else
  \placefigure{fig-vz}
\fi

This effect can be nicely explained by the Galactic warp. The Sun roughly lies 
at the inner edge of the warp on the line of nodes such that nearby stars that 
participate in the warp have $v_z>0$. For such stars to enter our very local 
sample, they must be near the pericenter of their orbits, and hence have $v_y>0
$. To be more quantitative, let me model the height $z$ of the Galactic disk as
a function of galacto-centric distance $R$ and azimuth $\varphi$ (with $\varphi
=0$ at the Sun) by
\beq 	
	z = h(R)\,\sin(\varphi+\Omega_p t),
\eeq 
where $\Omega_p$ is the pattern speed with $\Omega_p>0$ meaning a retrograde 
motion w.r.t.\ the stellar orbits, as expected from almost all theoretical warp 
models. I take the height function
\beq
	h(R)=(R-r_w)^2/r_h,
\eeq
where $r_w$ is the edge of the flat part of the disk and $r_h$ then sets the 
amplitude of the warp. For a star with guiding center $R$, we expect from the
conservation of angular momentum
\beq	
	v_y = {R^2\over R_0}\Omega(R) - R_0 \Omega(R_0) - v_{\odot y}
\eeq 
and a mean vertical motion of $\bar{v}_z= v_z(R)-v_z(R_0)-v_{\odot z}$
with (using $\varphi= \Omega t$)
\beq
	v_z(R) = {\D z \over \D t} 
	       = \big(\Omega(R)+\Omega_p\big)\,h(R)\,\cos\varphi,
\eeq
where $\Omega(R)$ is the circular frequency and $\bvel_\odot=($10.0, 5.2, 
7.2)\kms\ is the solar motion w.r.t.\ the LSR as derived in \paI. The lines in 
Figure~\ref{fig-vz} have been computed assuming a slightly falling rotation 
curve of the form $R\Omega\propto R^{-1/9}$, $R_0=8\kpc$, and $\Omega(R_0)=27.2
\kmskpc$ (\cite{fw97}). The dashed line is for a model with $r_w= 6.5\kpc$, 
$r_h=15\kpc$, given for the \hi\ by Diplas \& Savage (1990), and $\Omega_p=0$. 
Note that, for the rotation curve adopted, $v_y=40\kms$ originates from a 
guiding center of $R=10\kpc$. Clearly, this model does not fit the data and 
non-zero $\Omega_p$ would make it even worse%
\footnote{Of course, I made several simplifications, for instance, assuming that
	all stars are precisely at $R= R_0$. However, this is unlikely to 
	alter the conclusions. A more appropriate analysis, e.g.\ a Monte Carlo
	simulation, is beyond the scope of this paper.}.
The obvious reason for this failure is the fact that the modeled warp starts 
well inside the solar circle, and predicts a strong effect already near the LSR.
The full (almost indistinguishable) lines show four models with an inner edge 
of the warp of $r_w= R_0$, various pattern speeds in the range $0\le\Omega_p\le
30\kmskpc$, and with $r_h$ chosen such that the pattern speed and the amplitude
$z_{10}$ of the warp at $R=10\kpc$ obey
\beq\label{warp-ampl}
	{\Omega_p\over\kmskpc} = {4 - 6 (z_{10}/\kpc)\over 0.3 (z_{10}/\kpc)}.
\eeq
These models give good descriptions of the data at all valid points. Thus, to 
explain the observed effect, the Galactic warp must not start inside the solar 
circle. On the other hand, only a combination of the amplitude of the warp and 
its pattern speed is constraint by the data. With the above parameters for the
Galactic rotation curve and $R_0$, the \hi\ data of Burton (1992) yield a
warp amplitude $z_{10}$ of 0.3 to 0.4\kpc, resulting in $\Omega_p$ of
$13$ to $25\kmskpc$. The infrared data taken by DIRBE suggest that the stellar
warp might have half that amplitude (\cite{freu94}), in which case $\Omega_p$
must be larger.

\ifpreprint
  \section{CONCLUSION}
\else
  \section{Conclusion}
\fi
I have analyzed a kinematically unbiased sample of more than 14\,000 nearby
stars with positions, parallaxes, and proper motions known accurately from 
ESA's astrometric mission `\hip'. From the same sample, we have re-determined 
in \paI\ (\cite{paper1}), as a function of color, the mean velocity and 
velocity dispersion for main-sequence stars in the solar neighborhood. In
this paper, the velocity distribution $f(\bvel)$ itself, rather than its first 
two moments, has been inferred. 

Although radial velocities are available for a fraction of these stars, they 
could not be used because they are predominantly known for high-velocity stars 
and would inevitably introduce a kinematic bias. Without knowledge of radial 
velocities, the inference of $f(\bvel)$ is formally equivalent to a 
deprojection and an example of an astronomical inverse problem. A 
maximum-penalized-likelihood (MPL) technique for the recovery of $f(\bvel)$ from
the positions and tangential velocities of individual stars has been developed.

Poisson noise completely dominates the errors, and the number of stars is too 
small to recover the three-dimensional distribution with useful resolution, but
any two-dimensional projection can be determined with reasonable accuracy. The 
resulting distributions of nearby stars in $(v_x,v_y)$, $(v_x,v_z)$, and 
$(v_y,v_z)$ are shown in Figures~\ref{fig-fuv}, \ref{fig-fuw}, and 
\ref{fig-fvw}, respectively. ($v_x$, $v_y$, and $v_z$ denote the velocity 
components in the directions of the Galactic center, Galactic rotation, and the 
north Galactic pole, respectively). The sample has been split into four color 
bins (B1 to B4) of main-sequence stars and non-main-sequence stars (GI), mainly
giants. Additionally, the sample has been analyzed as a whole (AL).

\subsection{The Smooth Background}
At large velocities and/or redder color most stars are in a smooth and 
approximately ellipsoidal background distribution. The axis ratio of this 
background distribution increases from early to late stellar types. The 
increase is stronger in $v_z$ than it is in the horizontal motions, implying 
that the vertical heating becomes relatively more important for on average 
older and hence dynamically hotter stellar populations. This is expected if 
scattering by spiral structure is important, since for stars with epicycle 
diameters larger than the inter-arm separation, this process becomes inefficient
(\cite{jenkins}). From Figure~\ref{fig-fuw}, sample AL, I estimate that the 
effect sets in at $|v_x|$ of $\sim60\kms$ corresponding to an epicycle diameter 
of $\sim3\kpc$ (with $\kappa=36.7\kmskpc$).

The background distribution also shows nicely the asymmetric drift relation:
for ever later stellar types, the outermost contours in Figs.~\ref{fig-fuv} to
\ref{fig-fvw} are larger and their centroid is shifted more and more to
negative $v_y$. These contours appear to be aligned with the $v_x$ axis 
(radial) but slightly skewed w.r.t.\ the $v_y$ and $v_z$ axes. The
latter can nicely be explained by the Galactic warp: orbits participating in it
move upwards and must have $v_y>0$ in order to enter our local sample. A more
quantitative analysis reveals that the warp must not start within the solar
circle and has pattern speed $\gtrsim13\kmskpc$ with sense of rotation opposite
to the stellar orbits, as is theoretically expected.

\subsection{The Moving Groups}
Apart from the smooth background, $f(\bvel)$ shows a lot of structure, in
particular in the horizontal motions (Fig.~\ref{fig-fuv}): there are various
major and minor maxima and smaller features that seem to be real, in as much 
as they appear in more than one of the distinct subsamples. A peak in $f(\bvel)$
corresponds to a group of stars moving with the same velocity, and indeed, many
maxima correspond to one of the moving groups already identified by Kapteyn 
(1905) and later studied by Eggen (e.g.\ 1965). The fact that these moving 
groups are much less visible in $v_z$ can be attributed to phase-mixing. What 
seems to be surprising, however, is that the distribution of moving groups 
obeys an asymmetric drift relation, similar to the smooth background: the 
apparently older groups (as judged from their minimum color) are more wide 
spread in $(v_x,v_y)$ and lag w.r.t.\ the LSR, i.e.\ they are on non-circular 
orbits. A possible explanation for a moving groups on such an eccentric orbits 
is as follows. A cluster of stars is born onto a near-circular orbit similar 
to those of the molecular gas. The stars might then be trapped into a resonance
with a non-axisymmetric force field like that of the Galactic bar. If this 
force field slowly evolves, the resonances are shifted in orbit space and with 
them the stars trapped. This mechanism is well known from solar system dynamics
and has recently been proposed to explain the thick disk (\cite{sridtoum96}).

To interpret properly the results of this paper, i.e.\ to answer the question
why $f(\bvel)$ shows the observed structure, we clearly need a better 
understanding of the processes potentially involved in the dynamical evolution 
of the solar neighborhood, in particular, with regard to the effect of orbital 
resonances.  These processes include (i) phase-mixing; (ii) scattering by small
random perturbations of the Galactic force field, like giant molecular clouds 
and temporary spiral arms; (iii) interactions with large transient perturbers, 
for example, merging satellites of the Milky Way, such as the Saggritarius 
dwarf galaxy; and (iv) forcing by regular non-axisymmetric components of the 
gravitational potential, like the central bar, a long-living spiral pattern, a 
triaxial halo, or the Magellanic clouds. For the processes (i), (iii), and (iv),
resonances in the stellar motions and/or between these and the motions
of the perturbing agents are likely to be important and may lead to a behavior 
completely different from that of the non-resonant case.

%
%
\ifpreprint
  \section*{ACKNOWLEDGEMENTS}
\else
  \acknowledgements
\fi
I am grateful to James Binney for many discussions and Agris Kalnajs for useful
comments on an early version of this paper. Special thanks are due to David 
Merritt for pointing me to the work of Silverman and the technique of 
maximum-penalized-likelihood estimation.

%
%
\ifpreprint \else

\clearpage

\begin{deluxetable}{lrr@{\quad}r@{\quad}r@{\quad}c}
\small
\tablewidth{0pt}
\tablecaption{The subsamples analyzed \label{tab-s}}
\tablehead{
	\colhead{name} & 
	\multicolumn{2}{c}{$(\bv)_{\rm min,max}$} &
	\colhead{$N_{\rm tot}$} & \colhead{$N_{\rm in}$} &
	\colhead{$\tau_{\rm max}$}
}
\startdata
B1& - &0.0&  524&  524&$4\times10^8$ \nl
B2&0.0&0.4& 3201& 3199&$2\times10^9$ \nl
B3&0.4&0.6& 4596& 4582&$8\times10^9$ \nl
B4&0.6& - & 3544& 3527&      -       \nl
GI& - & - & 2504& 2491&      -       \nl
AL& - & - &14369&14323&      -
\enddata
\tablecomments{
Color limits (in mag), total number $N_{\rm tot}$ of stars, number 
$N_{\rm used}$ of stars used, and maximum stellar age (in years), for the color
bins B1 to B4 of main-sequence stars, the giant sample GI and the union of 
these, AL. (see text).
}
\end{deluxetable}

\clearpage

\begin{deluxetable}{r@{\quad}r@{\hspace{0.5em}}r@{\hspace{0.5em}}r@{\qquad}%
	c@{}c@{}c@{}c@{}c@{}c@{\qquad}l}
\small
\def\y{\multicolumn{1}{c}{$\bullet$}}
\def\q{\multicolumn{1}{c}{$\circ$}}
\tablewidth{0pt}
\tablecaption{Features in the inferred velocity distributions\label{tab-st}}
\tablehead{
	\colhead{No.} &
	$v_x$ & $v_y$ & $v_z$ &
	\colhead{B1}&\colhead{B2}&\colhead{B3}&\colhead{B4}&\colhead{GI} & &
	\colhead{moving group}
}
\startdata
 1&$-12$&$ -22$&$ -7$&\y&\y&\y&\y&\q&&Pleiades		\nl
 2&$-40$&$ -20$&$  0$&  &\y&\y&\y&\y&&Hyades		\nl
 3&$  9$&$   3$&$   $&\y&\y&\y&\y&\y&&Sirius \& UMa	\nl
 4&$-10$&$  -5$&$ -8$&\y&\y&\y&\y&  &&Coma Berenices	\nl
 5&$-25$&$ -10$&$-15$&  &\y&\y&\y&\q&&NGC\,1901		\nl
 6&$ 20$&$ -20$&$   $&  &  &  &\y&\y&&			\nl
 7&$ 15$&$ -60$&$   $&  &  &  &\y&\q&&HR1614		\nl
 8&$-40$&$ -50$&$   $&  &\y&\y&\y&\y&&			\nl
 9&$-25$&$ -50$&$   $&  &  &\q&\y&\y&&			\nl
10&$  0$&$-110$&$   $&\y&  &\q&\q&\q&&Arcturus		\nl
11&$-70$&$ -10$&$   $&  &  &\q&\q&\q&&			\nl
12&$-70$&$ -50$&$   $&  &  &\q&\q&\y&&			\nl
13&$ 50$&$   0$&$   $&  &  &\q&\y&\q&&			\nl
14&$ 50$&$ -25$&$   $&  &  &\q&\y&\q&&
\enddata
\tablecomments{Velocities are approximate ($v_x$ is $-U$ in Eggen's papers);
	if $v_z$ is missing it cannot be determined. $\bullet$ and $\circ$ 
	denote, respectively, clear or vague visibility in the corresponding 
	subsample. The list of associated moving groups is presumably 
	incomplete.}
\end{deluxetable}

\clearpage

\fi 

%
%
\ifpreprint
  \def\thebibliography#1{\subsection*{REFERENCES}
    \list{\null}{\leftmargin 1.2em\labelwidth0pt\labelsep0pt\itemindent -1.2em
    \itemsep0pt plus 0.1pt
    \parsep0pt plus 0.1pt
    \parskip0pt plus 0.1pt
    \usecounter{enumi}}
    \def\refpar{\relax}
    \def\newblock{\hskip .11em plus .33em minus .07em}
    \sloppy\clubpenalty4000\widowpenalty4000
    \sfcode`\.=1000\relax}
\fi

%
%

\ifpreprint \relax \else
\clearpage \onecolumn

\begin{figure}
\caption[]{Conjugate-gradient maximization of $\tcQ$ for a run with $L=96\times
	80\times56$ and $N=1294$ data drawn at random from a model for 
	$f(\bvel)$. The panels show as a function of the number of iterations 
	completed: (a) gradient of \tcQ, supremum and {\sc rms} value of the 
	increment; (b) integrated square error (\ref{eq-ise}); (c) 
	Kullback-Leiber information-distance (\ref{eq-kld}); and (d) \cQ\
	itself.}
        \label{fig-conv}\end{figure}

\begin{figure}\caption[]{
	The data: distribution in galactic coordinates of the five distinct
	sets B1 to B4 and GI. The bottom right frame shows lines of constant 
	$\ell$ (from $180^\circ$ at the left to $-180^\circ$ at the right, 
	spacing of $60^\circ$) and $b$ (from $-90^\circ$ at the bottom to 
	$90^\circ$ at the top, spacing of $30^\circ$), the triangles mark the 
	poles of the ecliptic.}
	\label{fig-sky}\end{figure}

\begin{figure}\caption[]{
	The distributions in $v_x$ (towards the Galactic center) and $v_y$
	(in direction of Galactic rotation): projection of $f(\bvel)$ obtained
	as MPLE for the sets listed in Table~\ref{tab-s}. Gray scales are linear
	and the contours contain, from inside outwards, 2, 6, 12, 21, 33, 50, 
	68, 80, 90, 95, 99, and 99.9 percent of all stars, i.e.\ half the 
	stars are within the innermost dark contour. The origin is at the 
	solar velocity, while the velocity derived for the LSR in \paI\ is 
	indicated by a triangle. Note that the smoothing is optimal for the 
	full sample (AL) only, while the results for the subsets are 
	under-smoothed. However, since the subsets are distinct, any feature 
	common to more than one of them is likely to be real.}
	\label{fig-fuv}\end{figure}

\begin{figure}\caption[]{
	The distributions in $v_x$ (towards the Galactic center) and $v_z$
	(towards the NGP). See Fig.~\ref{fig-fuv} for the contour levels, etc.}
	\label{fig-fuw}\end{figure}

\begin{figure}\caption[]{
	The distributions in $v_y$ (in direction of Galactic rotation) and $v_z$
	(towards the NGP). See Fig.~\ref{fig-fuv} for the contour levels, etc.}
	\label{fig-fvw}\end{figure}

\begin{figure}\caption[]{
	The mean vertical motion, $\bar{v}_z$, measured from sample AL as
	a function of of $v_y$. The errors bars show 1 $\sigma$ errors due to 
	Poisson noise; however, adjacent points are not independent but coupled
	by the smoothing. The triangle indicates the velocity of the LSR, and 
	the lines are derived from models for the Galactic warp, see text.}
	\label{fig-vz}\end{figure}

%
%

\clearpage

%
%
	\epsfxsize=10cm \epsfbox[20 155 325 710]{Dehnen.fig1.ps}
	\centerline{Fig.~\ref{fig-conv}}

%
%
\clearpage
	\epsfxsize=\textwidth \epsfbox[20 420 585 710]{Dehnen.fig2.ps}
	\centerline{Fig.~\ref{fig-sky}}

%
%
\clearpage
	\centerline{\epsfxsize=20.3pc\epsfbox[25 32 310 237]{Dehnen.fig3a.ps}
		    \epsfxsize=20.3pc\epsfbox[25 32 310 237]{Dehnen.fig3b.ps}}
	\centerline{\epsfxsize=20.3pc\epsfbox[25 32 310 237]{Dehnen.fig3c.ps}
		    \epsfxsize=20.3pc\epsfbox[25 32 310 237]{Dehnen.fig3d.ps}}
	\centerline{\epsfxsize=20.3pc\epsfbox[25 12 310 237]{Dehnen.fig3e.ps}
		    \epsfxsize=20.3pc\epsfbox[25 12 310 237]{Dehnen.fig3f.ps}}
	\centerline{ Fig.~\ref{fig-fuv}}

%
%
\clearpage
	\centerline{\epsfxsize=20.3pc\epsfbox[25 32 310 194]{Dehnen.fig4a.ps}
		    \epsfxsize=20.3pc\epsfbox[25 32 310 194]{Dehnen.fig4b.ps}}
	\centerline{\epsfxsize=20.3pc\epsfbox[25 32 310 194]{Dehnen.fig4c.ps}
		    \epsfxsize=20.3pc\epsfbox[25 32 310 194]{Dehnen.fig4d.ps}}
	\centerline{\epsfxsize=20.3pc\epsfbox[25 12 310 194]{Dehnen.fig4e.ps}
		    \epsfxsize=20.3pc\epsfbox[25 12 310 194]{Dehnen.fig4f.ps}}
	\centerline{ Fig.~\ref{fig-fuw}}

%
%
\clearpage
	\centerline{\epsfxsize=14.88pc\epsfbox[25 32 310 254]{Dehnen.fig5a.ps}
		    \epsfxsize=14.88pc\epsfbox[25 32 310 254]{Dehnen.fig5b.ps}}
	\centerline{\epsfxsize=14.88pc\epsfbox[25 32 310 254]{Dehnen.fig5c.ps}
		    \epsfxsize=14.88pc\epsfbox[25 32 310 254]{Dehnen.fig5d.ps}}
	\centerline{\epsfxsize=14.88pc\epsfbox[25 12 310 254]{Dehnen.fig5e.ps}
		    \epsfxsize=14.88pc\epsfbox[25 12 310 254]{Dehnen.fig5f.ps}}
	\centerline{ Fig.~\ref{fig-fvw}}
	\notetoeditor{
%
%
  Figs.~3, 4, and 5 come each in 6 files, one for each panel. The panels are 
  intended to be ordered in 2 columns and 3 rows. To see how this can be 
  achieved using \,{\tt epsf.sty}\, such that the tick-spacing is the same for 
  the three Figs., see lines 1328 to 1363 of file \,{\tt Dehnen.tex}.}
%
%

%
%
\clearpage
	\centerline{\epsfxsize=30pc \epsfbox[20 320 600 720]{Dehnen.fig6.ps}}
	\centerline{Fig.~\ref{fig-vz}}
\fi 
\end{document}